\documentclass[aps,prl,twocolumn,superscriptaddress,showpacs,showkeys]{revtex4-2}
\usepackage{graphicx}                   
\usepackage{hyperref}                   
\usepackage{amsmath,amssymb}            
\usepackage{epstopdf}
\usepackage{subcaption}					
\usepackage{float}
\usepackage{color}
\definecolor{orange}{rgb}{1,0.5,0}
\definecolor{brown}{rgb}{0.65, 0.16, 0.16}
\definecolor{phlox}{rgb}{0.87, 0.0, 1.0}

\graphicspath{{figs/}}					

\begin{document}

	\title{Stochastic and deterministic dynamics in networks with excitable nodes}
	
	\author{M. Rahimi-Majd}
	\affiliation{Department of Physics, Shahid Beheshti University, 1983969411, Tehran, Iran}
	\author{J. G. Restrepo}
	\affiliation{Department of Applied Mathematics, University of Colorado, Boulder, Colorado 80309, USA}
	\email{juanga@colorado.edu}
	\author{M. N. Najafi}
	\affiliation{Department of Physics, University of Mohaghegh Ardabili, P.O. Box 179, Ardabil, Iran}
	\email{morteza.nattagh@gmail.com}

\begin{abstract}
The analysis of the dynamics of a large class of excitable systems on  locally tree-like networks leads to the conclusion that at $\lambda=1$ a continuous phase transition takes place, where $\lambda$ is the largest eigenvalue of the adjacency matrix of the network. This paper is devoted to evaluate this claim for a more general case where the assumption of the linearity of the dynamical transfer function is violated with a non-linearity parameter $\beta$ which interpolates between stochastic ($\beta=0$) and deterministic ($\beta\rightarrow\infty$) dynamics. Our model shows a rich phase diagram with an absorbing state and extended critical and oscillatory regimes separated by transition and bifurcation lines which depend on the initial state. We test initial states with ($\mathbb{I}$) only one initial excited node, ($\mathbb{II}$) a fixed fraction  ($10\%$) of excited nodes, for all of which the transition is of first order for $\beta>0$ with a hysteresis effect and a gap function. For the case ($\mathbb{I}$) in the thermodynamic limit the absorbing state in the only phase for all $\lambda$ values and $\beta>0$. We further develop mean-field theories for  cases ($\mathbb{I}$) and ($\mathbb{II}$). For case ($\mathbb{II}$) we obtain an analytic  one-dimensional map which explains the essential properties of the model, including the hysteresis diagrams and fixed points of the dynamics.
\end{abstract}

	\pacs{05., 05.20.-y, 05.10.Ln, 05.45.Df}
	\keywords{complex networks, excitable nodes, first and second order transitions, deterministic and stochastic dynamics}
	\maketitle

The collective behavior of networks comprised of excitable nodes has found applications in many systems ranging from avalanches of neuronal bursting in the mammalian cortex~\cite{beggs2003neuronal,petermann2009spontaneous} and neuroscience as a whole~\cite{gerstner2014neuronal,stewart2008homeostasis,shew2009neuronal,larremore2012statistical,de2006self,de2010learning} to epidemiology~\cite{miller2009percolation,allard2009heterogeneous} and social systems~\cite{miller2009percolation,kinouchi2006optimal}, in all of which the effect of stochasticity in believed to be crucial. This stochasticity in the dynamics of the excitable nodes is understood in two levels: models with single excitable agents with stochastic external stimuli, and coarse-grained models where each node refers to many agents whose collective response to the stimuli is stochastic. This stochasticity was proven to be responsible for various behaviors of the system, like a second-order non-equilibrium phase transition separating the absorbing state from the super-critical state~\cite{larremore2011predicting,larremore2012statistical}, bifurcation and oscillatory behavior~\cite{moosavi2017refractory,najafi2019effect}, and a critical transition line with varying exponents~\cite{rahimi2021role}. Among a long list of various collective modes of the system, arguably criticality is believed to be the most important state affecting the system functioning, including optimal dynamical range~\cite{larremore2011predicting,larremore2011effects,kinouchi2006optimal,shew2009neuronal}, synaptic learning~\cite{de2010learning} and optimal information processing~\cite{kinouchi2006optimal}. This state can be achieved in a self-organized fashion (as expected in the brain) which is a source of many theoretical studies based on the network plasticity (brain plasticity in~\cite{de2006self}), and experimental studies, e.g., by tuning the ratios of excitation to inhibition for cortex slice cultures grown on planar microelectrode arrays~\cite{shew2009neuronal}. The basic assumption in the theoretical studies is the linearity of the master equations with respect to excitation probability of single node $p_t$~\cite{larremore2011predicting}. This assumption is however violated for systems with leading higher order terms with non-linear expansion around the transition point (see the following). The deterministic dynamics with threshold is an extreme example which applies for, e.g., sandpile-like dynamics on complex networks~\cite{najafi2014bak,najafi2020some,najafi2018statistical}. In this paper we uncover a rich phase space by introducing and analyzing an interpolation parameter $\beta$, which interpolates between stochastic and deterministic dynamics.\\

Working with the (stochastic) Kinouchi-Copelli (KC) model~\cite{kinouchi2006optimal}, it was shown by Larremore et al.~\cite{larremore2011predicting} that for networks with stochastic dynamics the largest eigenvalue $\lambda$ of the adjacency matrix $A$ (approximated to be $\lambda=\sigma\left\langle k \right\rangle $ for the Erd\H{o}s-R\'enyi network~\cite{restrepo2007approximating} where $\sigma$ is the maximum weight of links, and $\left\langle k\right\rangle$ is the average of the node degree) and its associated eigenvector play a prominent role in determining the functional form and consequently the status of the system, i.e., for $\lambda<\lambda_c\equiv 1$ the system is in the absorbing state where no node is excited, while for $\lambda>\lambda_c$ the activity of the system saturates eventually to a state where a macroscopic fraction of the nodes are ``on''. The key strategy in this work, that was followed later by many authors for other different dynamical models~\cite{larremore2011effects,larremore2012statistical,moosavi2017refractory} was linearizing the governing master equations and also the system's response function $F\equiv \left\langle f\right\rangle_t $ (with $\left\langle\right\rangle_t$ denoting an average over time and $f$ the fraction of excited nodes) with respect to the nodes' firing probability $p^t$ being small near the transition point $\lambda_c$. This strategy was proven to be general enough to give the properties of the system in the vicinity of the transition point given that the system response $F$ linearly vanishes when the external stimuli approaches zero~\cite{brochini2016phase}, which we call {\it linear stochasticity}. The inclusion of a refractory period causes additionally another transition point ($\lambda_b\equiv 2$) at which the system undergoes a bifurcation~\cite{moosavi2017refractory}. Therefore, the phase diagram of these systems is given by
\begin{equation}
\left\lbrace \begin{matrix}
0\le \lambda< \lambda_c & \text{subcritical regime}\\
\lambda_c< \lambda< \lambda_b & \ \ \ \ \ \ \ \ \text{extended critical regime}\\
\lambda>\lambda_b & \text{oscillatory regime}\\
\end{matrix}\right. .
\label{Eq:regimes}
\end{equation}
Many additional effects, like the impact of inhibitory nodes~\cite{larremore2014inhibition}, the effect of short-range sensory nodes (changing continuously $\lambda_c$ and $\lambda_b$)~\cite{rahimi2021role}, and retardation effects~\cite{najafi2019effect} were based on this analysis and the inspection of the activity-dependent branching ratio $b(M)$~\cite{martin2010activity}. \\

Here we show how the assumption of linear stochasticity can be relaxed (details in Appendix A). The network in the KC model~\cite{kinouchi2006optimal} is comprised of excitable nodes where each node $i$ can be in one of active or quiescent states, so that $x_i(t)=0 \;(1)$ characterizes the non-firing (firing) state, and $p_i^t=P(x_i(t)=1)$. Analytical treatments of this model usually assume locally tree-like networks which are defined as those networks where for $t$ not too large there is maximally one path of length $t$ for any pair $j$ and $k$. Besides the stochasticity in the dynamics, the basic assumption in~\cite{larremore2011predicting} is that \textit{one excited node} is enough to excite a neighboring node given that the later is not in the refractory period. Based on this, a master equation (Eq.~1 in SM) is developed and linearized (as well as $F$) with respect to $p^t$ and the input stimuli $x$ in the vicinity of the critical point $\lambda=\lambda_c$~\cite{larremore2011predicting,larremore2012statistical,larremore2011effects}. The violation of this linearity changes the thermodynamic properties of the system. Consider for example a locally tree network with $N$ excitable nodes in which \textit{$\alpha$ excited nodes} are required for exciting a destination node. Then, one can show that in terms of the KC model~\cite{kinouchi2006optimal} $F$ vanishes with the $\alpha$th power of $x$ in the limit $x\rightarrow 0$, and in the leading order (see Eq. 5 in SM)
\begin{equation}
p_i^{t+1}=\frac{1}{\alpha!}\left( \sum_{j=1}^Np_j^tA_{ij}\right)^{\alpha}
\end{equation} 
where $A_{ij}$ is the adjacency matrix of the network, i.e. $A_{ij}=w_{ij}$ if there is connection between $i$ and $j$ and zero otherwise, and $w_{ij}$ is a weight function. In an extreme limit $F$ is zero below a threshold and becomes non-zero abruptly beyond this threshold, which is a deterministic dynamics. Here we address this problem in a general setup using a parameter $\beta$ which besides making the dynamics non-linear, interpolates between linear stochastic ($\beta=0$) and nonlinear deterministic ($\beta=\infty$) dynamics, resulting in a rich phase diagram. In addition to studying the effect of a nonlinear transfer function, we explore the effect of a finite network. Theoretical analyses are usually carried out in the $N \to \infty$ limit, and initial conditions are chosen by implicitly assuming that a constant fraction of the nodes are initially excited. Here we also consider the case where $N$ is large but finite, and the system is started with only one excited node. We show that the effective phase diagram for this case is qualitatively different from the phase diagram obtained when a constant nonzero fraction of the $N \to \infty$ nodes is excited. 
\begin{figure}
	\includegraphics[width=80mm]{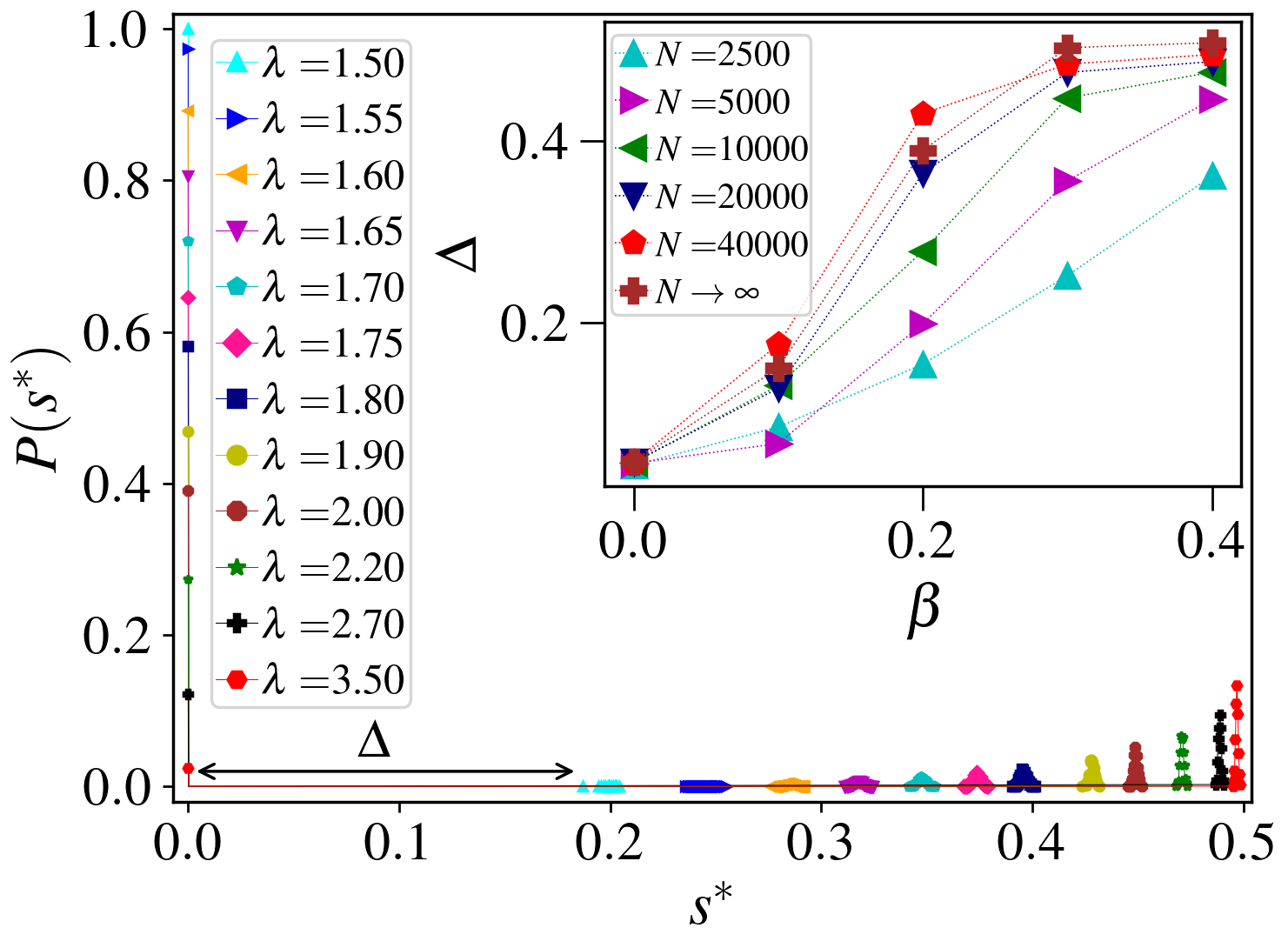}
	\caption{The distribution function of $s^*$ for various amounts of $\lambda$ for $\beta=0.2$ and $N=5000$ for the case $\mathbb{I}$.}
	\label{fig:p_x_star}
\end{figure}

Our model consists of $N$ excitable nodes on a random directed graph where each pair of nodes are connected with probability $q$, resulting in an average (equal) node in- and out-degree $\langle k \rangle=q(N-1)$. The connection weights $w_{i,j}$ are randomly distributed in the interval $[0,2\sigma]$, where $\sigma$ is a tuning parameter. The leading control parameter is the largest eigenvalue of $A_{ij}$, which is $\lambda\approx\sigma q N$ in our case (see Eq.~\ref{Eq:regimes}). The dynamics is like the KC model presented above, taking into account that some nodes are in the refractory period, i.e., the node cannot be excited immediately after being excited in the previous step. The probability that a node $i$ spikes at time $t+1$ is~\cite{larremore2014inhibition}
\begin{equation}
	\label{Eq1}
	p_i^{t+1}=\delta_{x_i(t),0} h_{\beta}\left( \sum_{j=1}^{N} A_{ij}x_j(t)\right) 
\end{equation}
where $\delta_{x_i(t),0}$ is unity if $x_i(t)=0$ and zero otherwise, i.e., it is the effect of the refractory period of one time step. $h_{\beta}$ is a dynamical transfer function which gives the probability that a node becomes active based on the total input, which is chosen to be $h_{\beta}(x)=G(h_0(x,\beta))$ where $G(x)$ is $x$ when $0\leq x\leq 1$, $1$ if $x>1$, and $0$ if $x < 0$, and $h_0(x,\beta)$ is
\begin{equation}
	h_0(x,\beta)=(2-\dfrac{2}{\pi}\tan^{-1}\beta)\dfrac{x^\beta}{x^\beta +1}G(x).
\end{equation}
The function $h_{\beta}(x)$ (Fig.~1 in SM) interpolates between linear stochastic ($\beta=0$ where the probability of spike increases linearly with the input potential~\cite{larremore2014inhibition,moosavi2017refractory}) and a step-like function of input which shows a deterministic dynamics ($\beta\rightarrow\infty$ where a node is excited only when the input potential exceeds a threshold, arbitrarily set to unity). When $\beta\rightarrow 0$, it was shown that in the continuum limit for large network sizes the model reduces to Wilson-Cowan equations~\cite{wilson1972excitatory} when both the inhibitory and excitatory nodes (neurons) are present~\cite{benayoun2010avalanches}. There it was shown that one way to reach criticality is adjusting the inhibitory and excitatory strengths in the system, which serve as a mechanism for neuronal avalanches~\cite{benayoun2010avalanches}.\\
As discussed above, in order to highlight the interplay between the initial conditions for the dynamics and finite-size effects in the system, two initial states (at $t=0$) are considered here: ($\mathbb{I}$) only one node is excited, and ($\mathbb{II}$) a macroscopic fraction of nodes (here $10\%$) are initially excited. For the case ($\mathbb{I}$) the transition between absorbing and supercritical phases is shown to be continuous for $\beta=0$ and is of first order for large $\beta$s, for which hysteretic behavior is observed. Our analytical and numerical results show that in the thermodynamic limit for $\beta>0$, the only dominant phase for the case ($\mathbb{I}$) is the absorbing state, i.e. initial conditions with $x^0$ invariable are attracted to $x^*=0$. For the case ($\mathbb{II}$) the transition lines are $\beta$-dependent in the thermodynamic limit. In particular, the bifurcation line jumps to $\lambda_b=4$ for non-zero $\beta$'s. Based on a mean field (MF) analysis we propose a one-dimensional map (by reducing Eq.~\ref{Eq1}) that gives us the behavior of the system in the thermodynamic limit of both cases, and enables us to treat the problem analytically. The MF theory is presented in the SM, where we developed MF theory for both cases separately, which is based on the analysis of the local activity defined as $s_t\equiv \frac{1}{N}\sum_{i=1}^Nx_i(t)$. Here we briefly summarize the approach for both cases. For case ($\mathbb{I}$) the main question is to determine when the excitation of a single node will lead to the absorbing state $s^* = 0$. One can easily show that when one node is initially excited, $x_n(0)=\delta_{n,i}$, the expected network activity at $t=1$ is given by
\begin{equation}
\mathbb{E}\left[s_1\right] =\frac{1}{N}\sum_{n\ne i}^N h_{\beta}\left( \sum_{m=1}^NA_{nm}x_m^0\right),
\end{equation}
where $\mathbb{E}\left[s_{t}\right]$ is the ensemble average of $s_{t}$. For an ER network with weight $w\in[0,2\sigma]$ one obtains (see SM)
\begin{equation}
	b= \frac{\mathbb{E}\left[s_1\right]}{s_0} = 	\left\langle k\right\rangle \int_0^1  h_{\beta}\left( \frac{2u\lambda}{\left\langle k\right\rangle }\right)du.
\end{equation}
One then can find the transition point by setting $b=1$: for $b < 1$ activity dies (on average) and approaches $s^* = 0$, and for $b > 1$ it becomes self-sustained. By expanding $h_{\beta}$ for the limit $N\rightarrow\infty$, one finds (see SM) that $\lambda_c\propto N^{\frac{\beta}{\beta+1}}$. More generally, the equation $b = 1$ gives the parameter curve separating $s^t \to 0$ or $s^t \to s_2 > 0$. \\
For case ($\mathbb{II}$), by assuming statistical independence of $\delta_{x_i^t,0}$ and $h_{\beta}\left( \sum_{j=1}^NA_{ij}x_j(t)\right) $ we have
\begin{equation}
	\mathbb{E}\left[s_{t+1}\right] =\left\langle \delta_{x_i^t,0}\right\rangle  \left\langle h_{\beta}\left( \sum_{j=1}^NA_{ij}x_j(t)\right)\right\rangle ,
\end{equation}
where $\mathbb{E}\left[s_{t+1}\right]$ is the ensemble average of $s_{t+1}$. Now we use the fact that $\left\langle \delta_{x_i^t,0}\right\rangle=1-s_t $. Furthermore, for an Erd\H{o}s-R\'enyi network with large mean degree the distribution of the random variable $\sum_{m=1}^NA_{nm}x_m^t$ is narrow about its mean $\lambda s_t$, so we approximate $\left\langle h_{\beta}\left( \sum_{j=1}^NA_{ij}x_j(t)\right)\right\rangle\simeq h_{\beta}\left(\lambda s_t \right) $, so that
\begin{equation}
\mathbb{E}\left[s_{t+1}\right] \simeq (1-s_t)h_{\beta}\left( \lambda s_t\right) 
\label{Eq:MF}
\end{equation}
By analyzing the dynamics of the one-dimensional map $s_{t+1} = (1-s_t) h_{\beta}(\lambda s_t)$ we are able to understand the observed behavior in case ($\mathbb{II}$).\\
\begin{figure}
	\includegraphics[width=80mm]{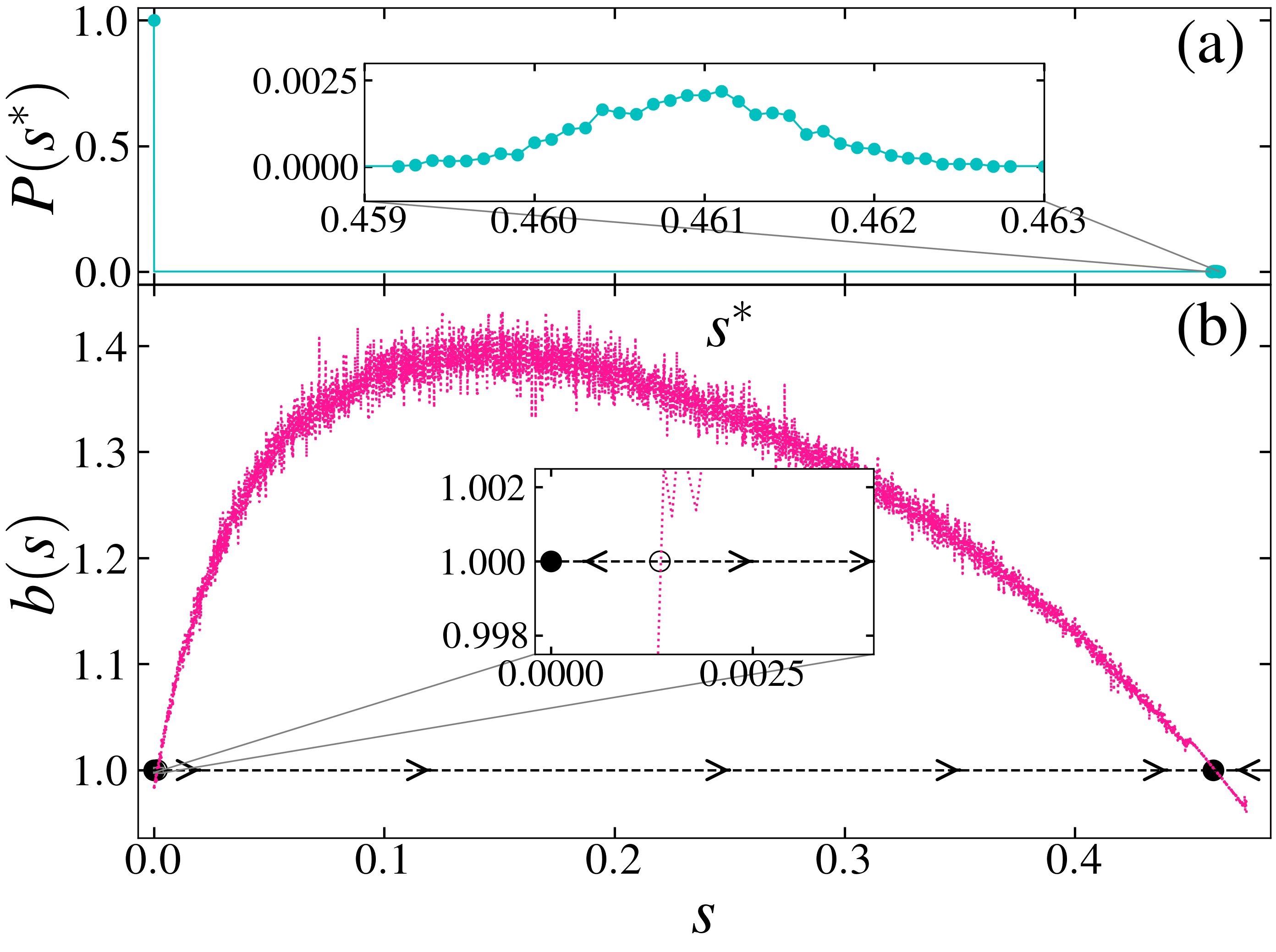}	
	\caption{ (a) The activity-dependent branching ratio. (a) $P(s^*)$ when the second peak is born for the first time. (b) the analysis for the branching ratio, showing the same structure.}
	\label{fig:b}
\end{figure}

\begin{figure*}
	\includegraphics[width=70mm]{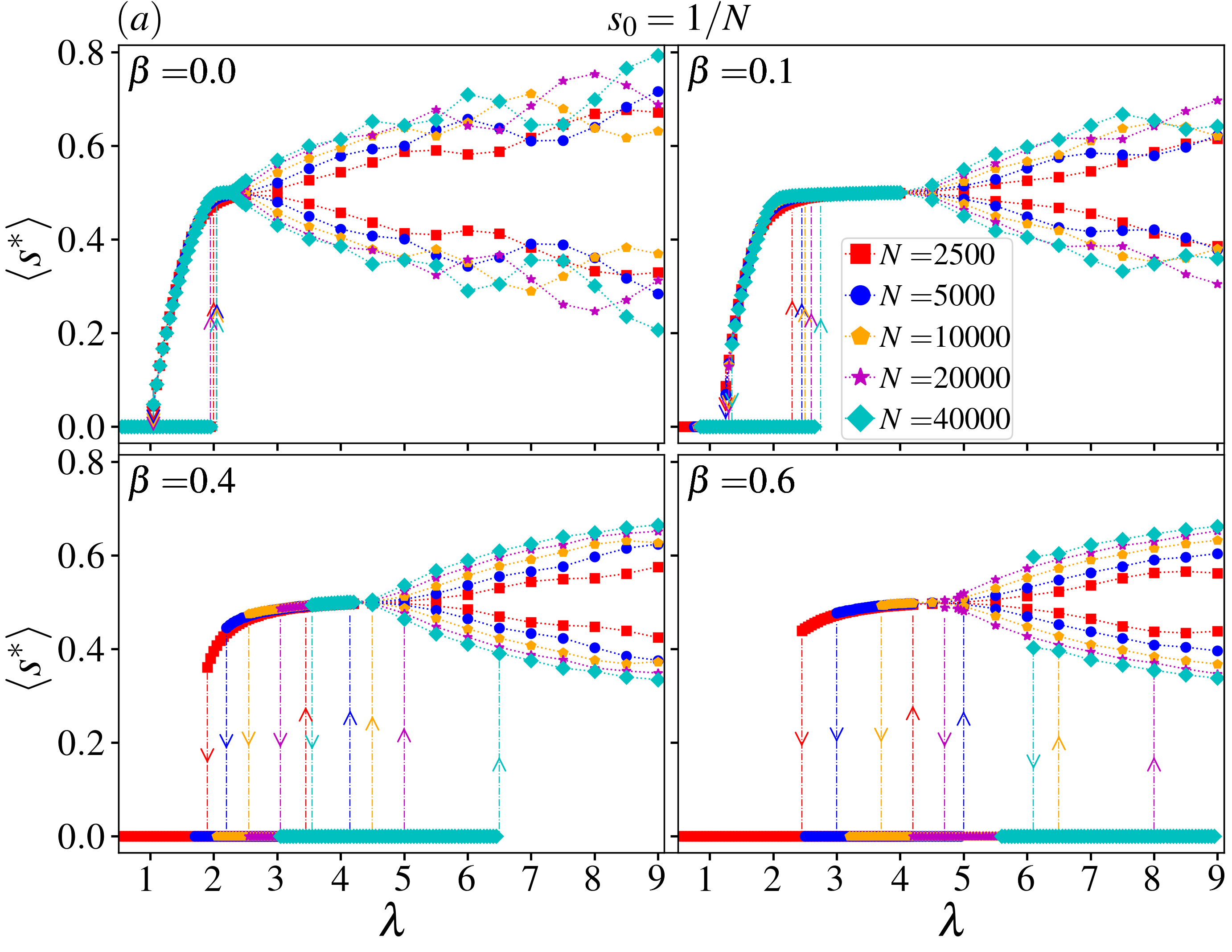}
	\includegraphics[width=70mm]{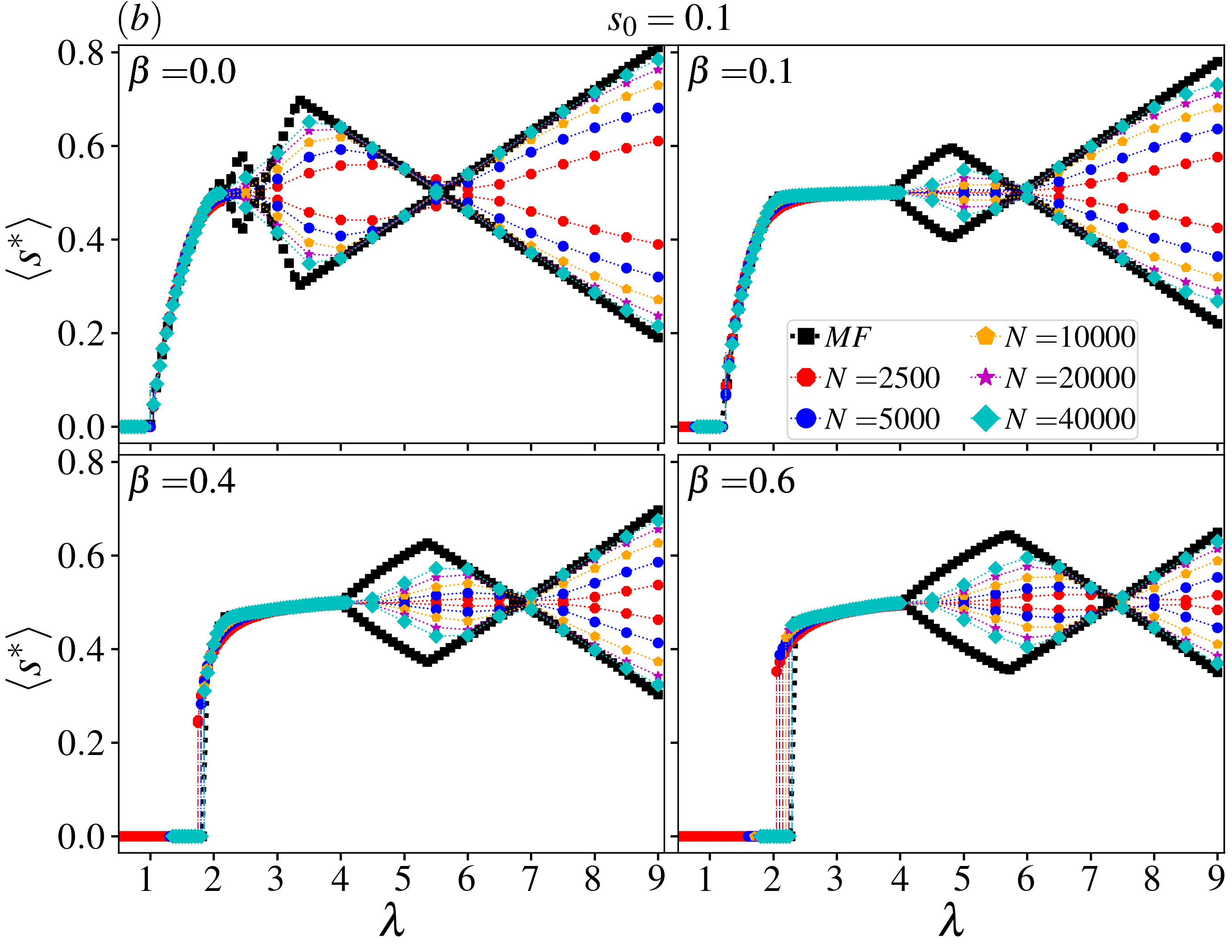}
	\caption{$\left\langle s^*\right\rangle $ in terms of $\lambda$ for various amounts of $N$ and $\beta$ values. The lines showing the hysteresis for two branches are identified by arrows. The left (a) and right (b) panels show the results for the case $\mathbb{I}$ and $\mathbb{II}$ respectively. In (b) the black bold square are the MF predictions. }
	\label{fig:x_star}
\end{figure*}
For the simulations we considered $\frac{N}{10^3}=2.5,5,10,22$ and $40$, for both ($\mathbb{I}$) and ($\mathbb{II}$) initial states ($x_i=0,1$ for quiescent and excited nodes respectively where $i$ is randomly chosen). $10^6$ samples were generated for each $\lambda$ and $\beta$ and $N$, over which the ensemble averages were taken. Let us consider $s^*$ as the fixed point of the dynamics, which is defined as $s^*\equiv\lim_{t\rightarrow\infty}\left\langle s_t\right\rangle_t$, so that $s^*=0$ for the absorbing state (subcritical regime), and is non-zero for extended critical regime. In the oscillatory phase $s$ oscillates between two non-zero limits. The probability distribution function (PDF) of $s^*$ for case ($\mathbb{I}$) is shown in Fig.~\ref{fig:p_x_star} in terms of $\beta$ which shows a bimodal structure, i.e. there are two peaks at $s^*_1=0$ and $s^*_2>0$. For very small $\lambda$ values the second peak is absent and there is only one peak at $s^*=0$, showing that the system is in the absorbing state regime. When we increase $\lambda$ the second peak $s^*_2$ is born for the first time in a point which we show by $\lambda^{(1)}_c(\beta)$. As $\lambda$ is increased further the position of the first peak is fixed, while the second peak moves to the right, and at the same time the height of the first (second) peak decreases (increases), and eventually the first peak dies at a point that we call $\lambda^{(2)}_c(\beta)$. This reveals that the system exhibits a discontinuity at the point where the second peak is born ($\lambda^{(1)}_c(\beta)$) using of which we define a \textit{gap parameter} as $\Delta(\beta)\equiv s^*_2-s^*_1$. As is shown in the inset of Fig.~\ref{fig:p_x_star}, $\Delta(\beta)$ is saturated in large enough $\beta$s, and also $\lim_{\beta\rightarrow 0}\Delta=0$, i.e. the gap closes at $\beta=0$. The same behaviors are observed (not shown here) for the case ($\mathbb{II}$), with smaller gap values.  The presence of two peaks with a gap in between is a signature of \textit{first order transition}, while the zero gap in $\beta=0$ suggests similarities with the \textit{second order transition} as is well-established in the literature.\\

We investigate here the activity-dependent branching ratio defined as $b(S)\equiv \mathbb{E}\left[\frac{s_{t+1}}{S}|s_t=S\right] $ to reveal the structure of the model, where $\mathbb{E}[A|B]$ represents ensemble average of $A$ conditioned on $B$~\cite{martin2010activity}. The fixed points are identified by the relation $b(s^*)=1$, so that the continuous transition points are identified by a monotonically decreasing $b(S)$ with the condition $\lim_{S\rightarrow 0}b(S)=1$. This function (computed numerically) is shown in Fig.~\ref{fig:b} for the case $\mathbb{I}$ for $\beta=0.3$, $N=10000$ and $\lambda=2.15$ showing two stable fixed points $s^*=0$ and $s^*\simeq 0.4608$ and one unstable fixed point $s^*_{\text{unstable}}\simeq 0.013$, which is consistent with the lower graph where the new peak starts to form for the first time. For $\beta=0$ at $\lambda=1$, $b(S)$ is a monotonically decreasing function of $S$ with the property $\lim_{S\rightarrow 0}b(S)=1$ as expected~\cite{larremore2014inhibition,moosavi2017refractory,najafi2019effect}. As a standard approach for bimodal PDFs~\cite{xue2020swarming}, we divide the data at the valley point between $s^*_1$ and $s^*_2$ (with PDFs represented by $P_1(s^*)$ and $P_2(s^*)$ respectively), and average to find two $\left\langle s^*\right\rangle_i $, $i=1,2$ for two branches (this enables us find the structure of the transition, otherwise a structure like $\beta=0$ case is obtained, see Fig.~3 in the SM). The results are shown in Fig.~\ref{fig:x_star}a (for the case $\mathbb{I}$) and~\ref{fig:x_star}b (for the case $\mathbb{II}$). In these graphs the upper branch (which is calculated with respect to $P_2(s^*)$) is born for the first time at $(\lambda^{(1)}_c(\beta),\Delta(\beta))$ as explained above, and the lower branch (which is calculated with respect to $P_1(s^*)$) dies at the point $(\lambda_c^{(2)}(\beta),0)$. As $\lambda$ increases further, the graph passes a bifurcation point $\lambda_b(\beta)$ beyond which the upper branch splits into (and oscillates between) two branches, the distance between which increases by increasing $\lambda$ (first observed by Moosavi \textit{et. al.}~\cite{moosavi2017refractory}). The bifurcation points are identified using the Kurthosis analysis, which for the transition point deviates from the Gaussian distribution~\cite{rahimi2021role}, see Fig.~5 in the SM. From this behavior shown in Fig.~\ref{fig:x_star} one observes a hysteresis behavior, i.e., the loops that are extended from $\lambda^{(1)}_c$ to $\lambda^{(2)}_c$ between two branches, shown for four $\beta$ values for both $\mathbb{I}$ and $\mathbb{II}$ cases (for more graphs see SM). The range in which the hysteresis effect is observed is much larger and much more $N$-dependent for the case $\mathbb{I}$ with respect to $\mathbb{II}$. The survival of the lower branch for the case $\mathbb{I}$ (much more than the case $\mathbb{II}$) and consequently a higher hysteresis range is expected since the dynamic starts from one excited node for which the probability of turning off of the whole network is higher. Observe that as $\beta$ increases the hysteresis effect magnifies and $\lambda_c^{(1,2)}$ grow with $\beta$ and $N$. The MF prediction (from iteration of the map $s_{t+1} = (1-s_t) h_{\beta}(\lambda s_t)$) is also shown in Fig.~\ref{fig:x_star}b with a bold continuous line to which the graph approaches asymptotically as $N\rightarrow\infty$. This figure shows intermittent oscillatory regime, which is consistent with MF as well as the kurtosis analysis (SM).\\

\begin{figure*}
	\includegraphics[width=85mm]{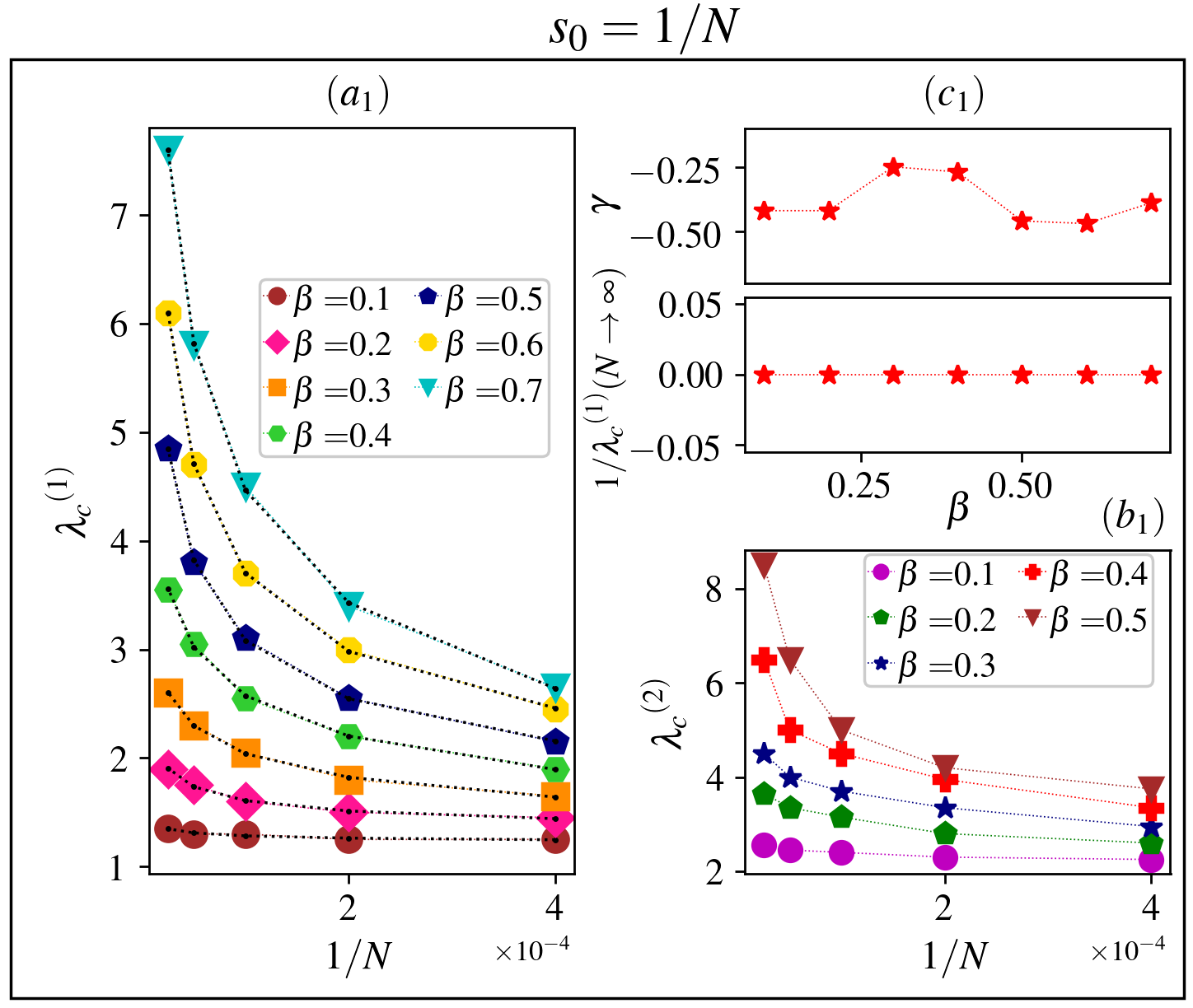}
	\includegraphics[width=85mm]{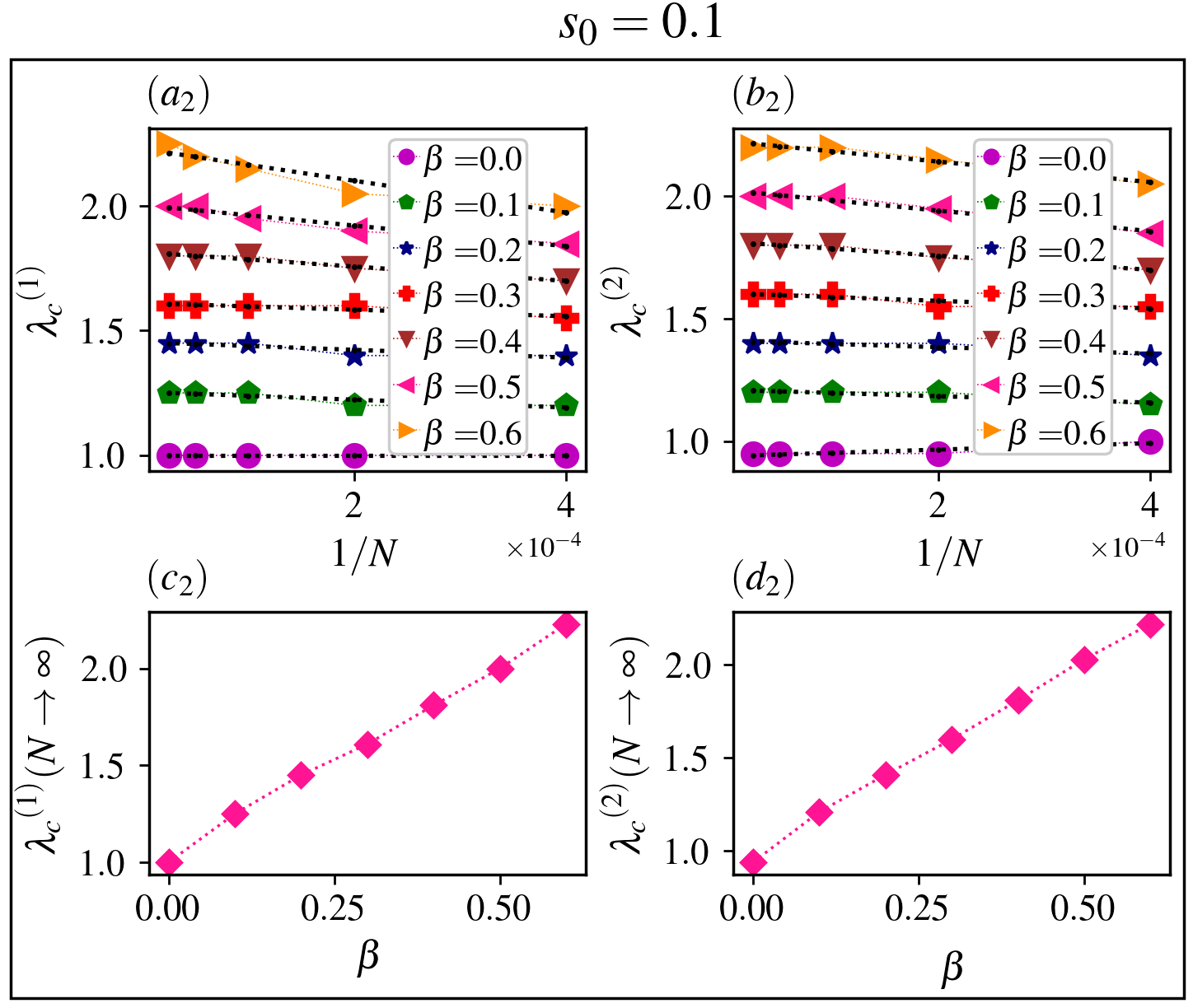}
	\caption{Left panel: results for the case $\mathbb{I}$, (a1) $\lambda^{(1)}_c$, (b1) $\lambda^{(2)}_c$ in terms of $1/N$ for which the fits are according to Eq.~\ref{Eq:gamma}. (c1) the exponent $\gamma$ and $1/\lambda_c^{(1)}$ in terms of $\beta$ for $N\rightarrow\infty$. Right panel: the same as left panel, for case $\mathbb{II}$}
	\label{fig:lambda}
\end{figure*}
For obtaining the behavior of the model in the thermodynamic limit, we traced that behavior of the transition points for various $N$s for both $\mathbb{I}$ and $\mathbb{II}$ cases. For the case $\mathbb{I}$ we found that the following fitting equation applies
\begin{equation}
	\lambda_c^{(1,2)}(N)=\lambda_c^{(1,2)}(\infty)+aN^{\gamma_{1,2}}
	\label{Eq:gamma}
\end{equation}
which is shown in Figs~\ref{fig:lambda}a1 and b1. The $\gamma$ exponent is reported in the Fig.~\ref{fig:lambda}c1 for $\beta>0$ (note that very small $\beta$ values are indistinguishable from $\beta=0$, and therefore we only have shown the results for $\beta\ge 0.1$). This reveals that for $\beta>0$ $\lambda_c^{(1,2)}(N\rightarrow\infty)\rightarrow\infty$, which is compatible with the MF prediction (the discrepancy with the exponent $\frac{\beta}{\beta+1}$ is due to the fact that the branching ratio test is for the critical point, while in $\lambda_c^{(1)}$ our system undergoes a discontinuous transition). We show $\lim_{N\rightarrow\infty}1/\lambda_c^{(1)}$ in this figure, which is almost zero for $\beta>0$ (the same holds for $\lambda_c^{(2)}$), confirming this claim. This observation implies that for $\beta>0$ the dominant phase is the absorbing state with stable fixed point $s^*=0$. For the case $\mathbb{II}$ the behavior is much more smooth and all the extrapolations of $\lambda_c^{(1,2)}$ are finite, see Figs.~\ref{fig:lambda}a2, b2, c2, and d2, see also SM. \\
\begin{figure}
	\includegraphics[width=90mm]{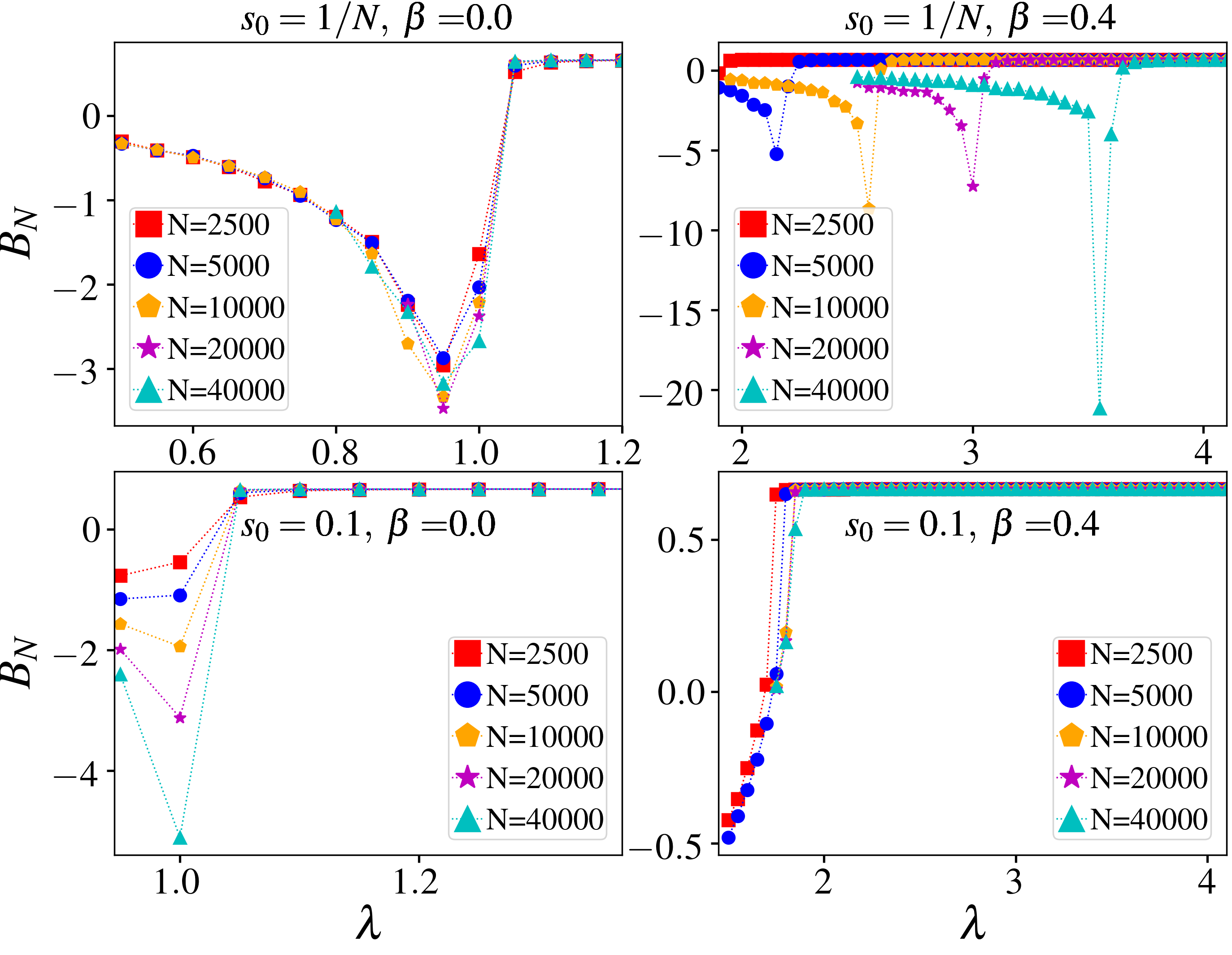}
	\caption{The Binder cumulant analysis for various $\beta$ and $N$ values.}
	\label{fig:binder}
\end{figure}
An interesting point for the case $\mathbb{I}$ is the structure of the model for $\beta=0$, for which the transition point is believed to be of second order~\cite{larremore2011effects,larremore2011predicting,larremore2014inhibition,moosavi2017refractory,najafi2019effect,rahimi2021role}. Figure~\ref{fig:x_star}a shows that for the transition in $\beta=0$, although the transition is gapless ($\Delta(\beta=0)=0$, see the inset in Fig.~\ref{fig:p_x_star}) there is hysteresis, i.e. $\lambda_c^{(2)}\ne\lambda_c^{(1)}\simeq 1$, showing that the transition point has similarities with the first order transitions. To test this more precisely, we consider the Binder cumulant defined as 
\begin{equation}
B_N= 1 - \frac{\left\langle s^4\right\rangle_N }{3\left\langle s^2\right\rangle^2_N }.
\end{equation}
This function enables us to distinguish first and second order phase transitions, such that it is continuous in second order transitions, while it shows a very sharp discontinuity for first order phase transitions, see for example~\cite{xue2020swarming}. We see this function for two $\beta$ values in Fig.~\ref{fig:binder} including $\beta=0$ (first column), which shows a sharp discontinuity exactly at the transition point, the strength of which increases with $N$. Now we discuss how the observed semi-continuous nature of the transition in case $\mathbb{I}$ can be reconciled with existing theoretical predictions and the continuous transition observed in in case $\mathbb{II}$. Note that the two peaks of the distribution of $s^*$ that lead to the zero and nonzero coexisting branches in the top left panel of Fig.~\ref{fig:x_star}a correspond, respectively, to realizations where the single initially excited node either fails or succeeds in generating self-sustained activity. Excitations of a single node failing to propagate in the supercritical regime is a feature that can not be captured by deterministic, mean field analyses done in the limit $N \to \infty$, and thus we view or results as complementary to these previous analyses.

\begin{figure}
	\includegraphics[width=80mm]{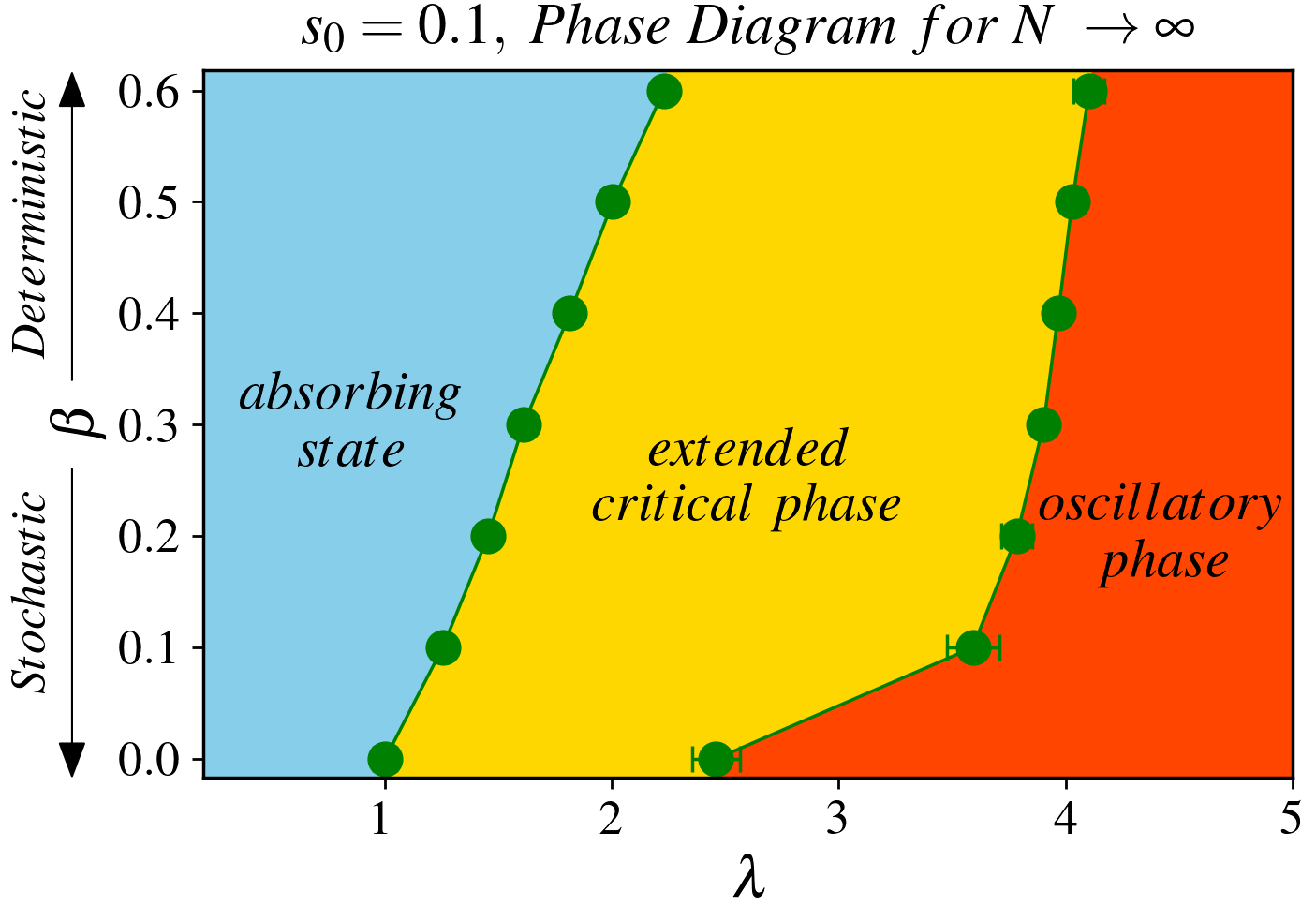}
	\caption{The phase diagram of the model for the case $\mathbb{II}$.}
	\label{fig:PhaseDiagram}
\end{figure}
To conclude, we investigated an interpolated model between linear-stochastic dynamics and deterministic dynamics. Numerical simulations, validated with a mean field analysis, show a rich phase diagram including hysteretic and period-2 dynamics in the thermodynamic limit. The phase diagram of the model is presented in Fig.~\ref{fig:PhaseDiagram} for case $\mathbb{II}$ (for the phase diagram for case $\mathbb{I}$ see SM). We have shown that the transitions for both $\mathbb{I}$ and $\mathbb{II}$ cases is of first order for $\beta>0$ for which some hysteresis effects were observed. For case $\mathbb{I}$ in the thermodynamic limit for $\beta>0$ the dominant phase is the one for the absorbing state, i.e. $s^*=0$ is a stable fixed point of the dynamics. For case $\mathbb{II}$, however, we observed that all sub-critical, super-critical and oscillating phases are present in the thermodynamic limit. Using MF arguments, we found a one-dimensional map which gives the properties of the model in the thermodynamic limit for both cases $\mathbb{I}$ and $\mathbb{II}$. Our results generalize previous analyses to the common case of a nonlinear transfer function and highlight the importance of stochastic effects when considering a small set of initially excited nodes.

-----------------------

\bibliography{refs}

\begin{thebibliography}{26}%
\makeatletter
\providecommand \@ifxundefined [1]{%
 \@ifx{#1\undefined}
}%
\providecommand \@ifnum [1]{%
 \ifnum #1\expandafter \@firstoftwo
 \else \expandafter \@secondoftwo
 \fi
}%
\providecommand \@ifx [1]{%
 \ifx #1\expandafter \@firstoftwo
 \else \expandafter \@secondoftwo
 \fi
}%
\providecommand \natexlab [1]{#1}%
\providecommand \enquote  [1]{``#1''}%
\providecommand \bibnamefont  [1]{#1}%
\providecommand \bibfnamefont [1]{#1}%
\providecommand \citenamefont [1]{#1}%
\providecommand \href@noop [0]{\@secondoftwo}%
\providecommand \href [0]{\begingroup \@sanitize@url \@href}%
\providecommand \@href[1]{\@@startlink{#1}\@@href}%
\providecommand \@@href[1]{\endgroup#1\@@endlink}%
\providecommand \@sanitize@url [0]{\catcode `\\12\catcode `\$12\catcode
  `\&12\catcode `\#12\catcode `\^12\catcode `\_12\catcode `\%12\relax}%
\providecommand \@@startlink[1]{}%
\providecommand \@@endlink[0]{}%
\providecommand \url  [0]{\begingroup\@sanitize@url \@url }%
\providecommand \@url [1]{\endgroup\@href {#1}{\urlprefix }}%
\providecommand \urlprefix  [0]{URL }%
\providecommand \Eprint [0]{\href }%
\providecommand \doibase [0]{http://dx.doi.org/}%
\providecommand \selectlanguage [0]{\@gobble}%
\providecommand \bibinfo  [0]{\@secondoftwo}%
\providecommand \bibfield  [0]{\@secondoftwo}%
\providecommand \translation [1]{[#1]}%
\providecommand \BibitemOpen [0]{}%
\providecommand \bibitemStop [0]{}%
\providecommand \bibitemNoStop [0]{.\EOS\space}%
\providecommand \EOS [0]{\spacefactor3000\relax}%
\providecommand \BibitemShut  [1]{\csname bibitem#1\endcsname}%
\let\auto@bib@innerbib\@empty
\bibitem [{\citenamefont {Beggs}\ and\ \citenamefont
  {Plenz}(2003)}]{beggs2003neuronal}%
  \BibitemOpen
  \bibfield  {author} {\bibinfo {author} {\bibfnamefont {J.~M.}\ \bibnamefont
  {Beggs}}\ and\ \bibinfo {author} {\bibfnamefont {D.}~\bibnamefont {Plenz}},\
  }\href@noop {} {\bibfield  {journal} {\bibinfo  {journal} {Journal of
  neuroscience}\ }\textbf {\bibinfo {volume} {23}},\ \bibinfo {pages} {11167}
  (\bibinfo {year} {2003})}\BibitemShut {NoStop}%
\bibitem [{\citenamefont {Petermann}\ \emph {et~al.}(2009)\citenamefont
  {Petermann}, \citenamefont {Thiagarajan}, \citenamefont {Lebedev},
  \citenamefont {Nicolelis}, \citenamefont {Chialvo},\ and\ \citenamefont
  {Plenz}}]{petermann2009spontaneous}%
  \BibitemOpen
  \bibfield  {author} {\bibinfo {author} {\bibfnamefont {T.}~\bibnamefont
  {Petermann}}, \bibinfo {author} {\bibfnamefont {T.~C.}\ \bibnamefont
  {Thiagarajan}}, \bibinfo {author} {\bibfnamefont {M.~A.}\ \bibnamefont
  {Lebedev}}, \bibinfo {author} {\bibfnamefont {M.~A.}\ \bibnamefont
  {Nicolelis}}, \bibinfo {author} {\bibfnamefont {D.~R.}\ \bibnamefont
  {Chialvo}}, \ and\ \bibinfo {author} {\bibfnamefont {D.}~\bibnamefont
  {Plenz}},\ }\href@noop {} {\bibfield  {journal} {\bibinfo  {journal}
  {Proceedings of the National Academy of Sciences}\ }\textbf {\bibinfo
  {volume} {106}},\ \bibinfo {pages} {15921} (\bibinfo {year}
  {2009})}\BibitemShut {NoStop}%
\bibitem [{\citenamefont {Gerstner}\ \emph {et~al.}(2014)\citenamefont
  {Gerstner}, \citenamefont {Kistler}, \citenamefont {Naud},\ and\
  \citenamefont {Paninski}}]{gerstner2014neuronal}%
  \BibitemOpen
  \bibfield  {author} {\bibinfo {author} {\bibfnamefont {W.}~\bibnamefont
  {Gerstner}}, \bibinfo {author} {\bibfnamefont {W.~M.}\ \bibnamefont
  {Kistler}}, \bibinfo {author} {\bibfnamefont {R.}~\bibnamefont {Naud}}, \
  and\ \bibinfo {author} {\bibfnamefont {L.}~\bibnamefont {Paninski}},\
  }\href@noop {} {\emph {\bibinfo {title} {Neuronal dynamics: From single
  neurons to networks and models of cognition}}}\ (\bibinfo  {publisher}
  {Cambridge University Press},\ \bibinfo {year} {2014})\BibitemShut {NoStop}%
\bibitem [{\citenamefont {Stewart}\ and\ \citenamefont
  {Plenz}(2008)}]{stewart2008homeostasis}%
  \BibitemOpen
  \bibfield  {author} {\bibinfo {author} {\bibfnamefont {C.~V.}\ \bibnamefont
  {Stewart}}\ and\ \bibinfo {author} {\bibfnamefont {D.}~\bibnamefont
  {Plenz}},\ }\href@noop {} {\bibfield  {journal} {\bibinfo  {journal} {Journal
  of neuroscience methods}\ }\textbf {\bibinfo {volume} {169}},\ \bibinfo
  {pages} {405} (\bibinfo {year} {2008})}\BibitemShut {NoStop}%
\bibitem [{\citenamefont {Shew}\ \emph {et~al.}(2009)\citenamefont {Shew},
  \citenamefont {Yang}, \citenamefont {Petermann}, \citenamefont {Roy},\ and\
  \citenamefont {Plenz}}]{shew2009neuronal}%
  \BibitemOpen
  \bibfield  {author} {\bibinfo {author} {\bibfnamefont {W.~L.}\ \bibnamefont
  {Shew}}, \bibinfo {author} {\bibfnamefont {H.}~\bibnamefont {Yang}}, \bibinfo
  {author} {\bibfnamefont {T.}~\bibnamefont {Petermann}}, \bibinfo {author}
  {\bibfnamefont {R.}~\bibnamefont {Roy}}, \ and\ \bibinfo {author}
  {\bibfnamefont {D.}~\bibnamefont {Plenz}},\ }\href@noop {} {\bibfield
  {journal} {\bibinfo  {journal} {Journal of neuroscience}\ }\textbf {\bibinfo
  {volume} {29}},\ \bibinfo {pages} {15595} (\bibinfo {year}
  {2009})}\BibitemShut {NoStop}%
\bibitem [{\citenamefont {Larremore}\ \emph {et~al.}(2012)\citenamefont
  {Larremore}, \citenamefont {Carpenter}, \citenamefont {Ott},\ and\
  \citenamefont {Restrepo}}]{larremore2012statistical}%
  \BibitemOpen
  \bibfield  {author} {\bibinfo {author} {\bibfnamefont {D.~B.}\ \bibnamefont
  {Larremore}}, \bibinfo {author} {\bibfnamefont {M.~Y.}\ \bibnamefont
  {Carpenter}}, \bibinfo {author} {\bibfnamefont {E.}~\bibnamefont {Ott}}, \
  and\ \bibinfo {author} {\bibfnamefont {J.~G.}\ \bibnamefont {Restrepo}},\
  }\href@noop {} {\bibfield  {journal} {\bibinfo  {journal} {Physical Review
  E}\ }\textbf {\bibinfo {volume} {85}},\ \bibinfo {pages} {066131} (\bibinfo
  {year} {2012})}\BibitemShut {NoStop}%
\bibitem [{\citenamefont {De~Arcangelis}\ \emph {et~al.}(2006)\citenamefont
  {De~Arcangelis}, \citenamefont {Perrone-Capano},\ and\ \citenamefont
  {Herrmann}}]{de2006self}%
  \BibitemOpen
  \bibfield  {author} {\bibinfo {author} {\bibfnamefont {L.}~\bibnamefont
  {De~Arcangelis}}, \bibinfo {author} {\bibfnamefont {C.}~\bibnamefont
  {Perrone-Capano}}, \ and\ \bibinfo {author} {\bibfnamefont {H.~J.}\
  \bibnamefont {Herrmann}},\ }\href@noop {} {\bibfield  {journal} {\bibinfo
  {journal} {Physical review letters}\ }\textbf {\bibinfo {volume} {96}},\
  \bibinfo {pages} {028107} (\bibinfo {year} {2006})}\BibitemShut {NoStop}%
\bibitem [{\citenamefont {de~Arcangelis}\ and\ \citenamefont
  {Herrmann}(2010)}]{de2010learning}%
  \BibitemOpen
  \bibfield  {author} {\bibinfo {author} {\bibfnamefont {L.}~\bibnamefont
  {de~Arcangelis}}\ and\ \bibinfo {author} {\bibfnamefont {H.~J.}\ \bibnamefont
  {Herrmann}},\ }\href@noop {} {\bibfield  {journal} {\bibinfo  {journal}
  {Proceedings of the National Academy of Sciences}\ }\textbf {\bibinfo
  {volume} {107}},\ \bibinfo {pages} {3977} (\bibinfo {year}
  {2010})}\BibitemShut {NoStop}%
\bibitem [{\citenamefont {Miller}(2009)}]{miller2009percolation}%
  \BibitemOpen
  \bibfield  {author} {\bibinfo {author} {\bibfnamefont {J.~C.}\ \bibnamefont
  {Miller}},\ }\href@noop {} {\bibfield  {journal} {\bibinfo  {journal}
  {Physical Review E}\ }\textbf {\bibinfo {volume} {80}},\ \bibinfo {pages}
  {020901} (\bibinfo {year} {2009})}\BibitemShut {NoStop}%
\bibitem [{\citenamefont {Allard}\ \emph {et~al.}(2009)\citenamefont {Allard},
  \citenamefont {No{\"e}l}, \citenamefont {Dub{\'e}},\ and\ \citenamefont
  {Pourbohloul}}]{allard2009heterogeneous}%
  \BibitemOpen
  \bibfield  {author} {\bibinfo {author} {\bibfnamefont {A.}~\bibnamefont
  {Allard}}, \bibinfo {author} {\bibfnamefont {P.-A.}\ \bibnamefont
  {No{\"e}l}}, \bibinfo {author} {\bibfnamefont {L.~J.}\ \bibnamefont
  {Dub{\'e}}}, \ and\ \bibinfo {author} {\bibfnamefont {B.}~\bibnamefont
  {Pourbohloul}},\ }\href@noop {} {\bibfield  {journal} {\bibinfo  {journal}
  {Physical Review E}\ }\textbf {\bibinfo {volume} {79}},\ \bibinfo {pages}
  {036113} (\bibinfo {year} {2009})}\BibitemShut {NoStop}%
\bibitem [{\citenamefont {Kinouchi}\ and\ \citenamefont
  {Copelli}(2006)}]{kinouchi2006optimal}%
  \BibitemOpen
  \bibfield  {author} {\bibinfo {author} {\bibfnamefont {O.}~\bibnamefont
  {Kinouchi}}\ and\ \bibinfo {author} {\bibfnamefont {M.}~\bibnamefont
  {Copelli}},\ }\href@noop {} {\bibfield  {journal} {\bibinfo  {journal}
  {Nature physics}\ }\textbf {\bibinfo {volume} {2}},\ \bibinfo {pages} {348}
  (\bibinfo {year} {2006})}\BibitemShut {NoStop}%
\bibitem [{\citenamefont {Larremore}\ \emph
  {et~al.}(2011{\natexlab{a}})\citenamefont {Larremore}, \citenamefont {Shew},\
  and\ \citenamefont {Restrepo}}]{larremore2011predicting}%
  \BibitemOpen
  \bibfield  {author} {\bibinfo {author} {\bibfnamefont {D.~B.}\ \bibnamefont
  {Larremore}}, \bibinfo {author} {\bibfnamefont {W.~L.}\ \bibnamefont {Shew}},
  \ and\ \bibinfo {author} {\bibfnamefont {J.~G.}\ \bibnamefont {Restrepo}},\
  }\href@noop {} {\bibfield  {journal} {\bibinfo  {journal} {Physical review
  letters}\ }\textbf {\bibinfo {volume} {106}},\ \bibinfo {pages} {058101}
  (\bibinfo {year} {2011}{\natexlab{a}})}\BibitemShut {NoStop}%
\bibitem [{\citenamefont {Moosavi}\ \emph {et~al.}(2017)\citenamefont
  {Moosavi}, \citenamefont {Montakhab},\ and\ \citenamefont
  {Valizadeh}}]{moosavi2017refractory}%
  \BibitemOpen
  \bibfield  {author} {\bibinfo {author} {\bibfnamefont {S.~A.}\ \bibnamefont
  {Moosavi}}, \bibinfo {author} {\bibfnamefont {A.}~\bibnamefont {Montakhab}},
  \ and\ \bibinfo {author} {\bibfnamefont {A.}~\bibnamefont {Valizadeh}},\
  }\href@noop {} {\bibfield  {journal} {\bibinfo  {journal} {Scientific
  reports}\ }\textbf {\bibinfo {volume} {7}},\ \bibinfo {pages} {1} (\bibinfo
  {year} {2017})}\BibitemShut {NoStop}%
\bibitem [{\citenamefont {Najafi}\ and\ \citenamefont
  {Rahimi-Majd}(2019)}]{najafi2019effect}%
  \BibitemOpen
  \bibfield  {author} {\bibinfo {author} {\bibfnamefont {M.}~\bibnamefont
  {Najafi}}\ and\ \bibinfo {author} {\bibfnamefont {M.}~\bibnamefont
  {Rahimi-Majd}},\ }\href@noop {} {\bibfield  {journal} {\bibinfo  {journal}
  {Physica Scripta}\ }\textbf {\bibinfo {volume} {94}},\ \bibinfo {pages}
  {055208} (\bibinfo {year} {2019})}\BibitemShut {NoStop}%
\bibitem [{\citenamefont {Rahimi-Majd}\ \emph {et~al.}(2021)\citenamefont
  {Rahimi-Majd}, \citenamefont {Seifi}, \citenamefont {de~Arcangelis},\ and\
  \citenamefont {Najafi}}]{rahimi2021role}%
  \BibitemOpen
  \bibfield  {author} {\bibinfo {author} {\bibfnamefont {M.}~\bibnamefont
  {Rahimi-Majd}}, \bibinfo {author} {\bibfnamefont {M.}~\bibnamefont {Seifi}},
  \bibinfo {author} {\bibfnamefont {L.}~\bibnamefont {de~Arcangelis}}, \ and\
  \bibinfo {author} {\bibfnamefont {M.}~\bibnamefont {Najafi}},\ }\href@noop {}
  {\bibfield  {journal} {\bibinfo  {journal} {Physical Review E}\ }\textbf
  {\bibinfo {volume} {103}},\ \bibinfo {pages} {042402} (\bibinfo {year}
  {2021})}\BibitemShut {NoStop}%
\bibitem [{\citenamefont {Larremore}\ \emph
  {et~al.}(2011{\natexlab{b}})\citenamefont {Larremore}, \citenamefont {Shew},
  \citenamefont {Ott},\ and\ \citenamefont {Restrepo}}]{larremore2011effects}%
  \BibitemOpen
  \bibfield  {author} {\bibinfo {author} {\bibfnamefont {D.~B.}\ \bibnamefont
  {Larremore}}, \bibinfo {author} {\bibfnamefont {W.~L.}\ \bibnamefont {Shew}},
  \bibinfo {author} {\bibfnamefont {E.}~\bibnamefont {Ott}}, \ and\ \bibinfo
  {author} {\bibfnamefont {J.~G.}\ \bibnamefont {Restrepo}},\ }\href@noop {}
  {\bibfield  {journal} {\bibinfo  {journal} {Chaos: An Interdisciplinary
  Journal of Nonlinear Science}\ }\textbf {\bibinfo {volume} {21}},\ \bibinfo
  {pages} {025117} (\bibinfo {year} {2011}{\natexlab{b}})}\BibitemShut
  {NoStop}%
\bibitem [{\citenamefont {Najafi}(2014)}]{najafi2014bak}%
  \BibitemOpen
  \bibfield  {author} {\bibinfo {author} {\bibfnamefont {M.}~\bibnamefont
  {Najafi}},\ }\href@noop {} {\bibfield  {journal} {\bibinfo  {journal}
  {Physics Letters A}\ }\textbf {\bibinfo {volume} {378}},\ \bibinfo {pages}
  {2008} (\bibinfo {year} {2014})}\BibitemShut {NoStop}%
\bibitem [{\citenamefont {Najafi}\ \emph {et~al.}(2020)\citenamefont {Najafi},
  \citenamefont {Tizdast},\ and\ \citenamefont
  {Cheraghalizadeh}}]{najafi2020some}%
  \BibitemOpen
  \bibfield  {author} {\bibinfo {author} {\bibfnamefont {M.}~\bibnamefont
  {Najafi}}, \bibinfo {author} {\bibfnamefont {S.}~\bibnamefont {Tizdast}}, \
  and\ \bibinfo {author} {\bibfnamefont {J.}~\bibnamefont {Cheraghalizadeh}},\
  }\href@noop {} {\bibfield  {journal} {\bibinfo  {journal} {arXiv preprint
  arXiv:2009.08160}\ } (\bibinfo {year} {2020})}\BibitemShut {NoStop}%
\bibitem [{\citenamefont {Najafi}\ and\ \citenamefont
  {Dashti-Naserabadi}(2018)}]{najafi2018statistical}%
  \BibitemOpen
  \bibfield  {author} {\bibinfo {author} {\bibfnamefont {M.}~\bibnamefont
  {Najafi}}\ and\ \bibinfo {author} {\bibfnamefont {H.}~\bibnamefont
  {Dashti-Naserabadi}},\ }\href@noop {} {\bibfield  {journal} {\bibinfo
  {journal} {Physical Review E}\ }\textbf {\bibinfo {volume} {97}},\ \bibinfo
  {pages} {032108} (\bibinfo {year} {2018})}\BibitemShut {NoStop}%
\bibitem [{\citenamefont {Restrepo}\ \emph {et~al.}(2007)\citenamefont
  {Restrepo}, \citenamefont {Ott},\ and\ \citenamefont
  {Hunt}}]{restrepo2007approximating}%
  \BibitemOpen
  \bibfield  {author} {\bibinfo {author} {\bibfnamefont {J.~G.}\ \bibnamefont
  {Restrepo}}, \bibinfo {author} {\bibfnamefont {E.}~\bibnamefont {Ott}}, \
  and\ \bibinfo {author} {\bibfnamefont {B.~R.}\ \bibnamefont {Hunt}},\
  }\href@noop {} {\bibfield  {journal} {\bibinfo  {journal} {Physical Review
  E}\ }\textbf {\bibinfo {volume} {76}},\ \bibinfo {pages} {056119} (\bibinfo
  {year} {2007})}\BibitemShut {NoStop}%
\bibitem [{\citenamefont {Brochini}\ \emph {et~al.}(2016)\citenamefont
  {Brochini}, \citenamefont {de~Andrade~Costa}, \citenamefont {Abadi},
  \citenamefont {Roque}, \citenamefont {Stolfi},\ and\ \citenamefont
  {Kinouchi}}]{brochini2016phase}%
  \BibitemOpen
  \bibfield  {author} {\bibinfo {author} {\bibfnamefont {L.}~\bibnamefont
  {Brochini}}, \bibinfo {author} {\bibfnamefont {A.}~\bibnamefont
  {de~Andrade~Costa}}, \bibinfo {author} {\bibfnamefont {M.}~\bibnamefont
  {Abadi}}, \bibinfo {author} {\bibfnamefont {A.~C.}\ \bibnamefont {Roque}},
  \bibinfo {author} {\bibfnamefont {J.}~\bibnamefont {Stolfi}}, \ and\ \bibinfo
  {author} {\bibfnamefont {O.}~\bibnamefont {Kinouchi}},\ }\href@noop {}
  {\bibfield  {journal} {\bibinfo  {journal} {Scientific reports}\ }\textbf
  {\bibinfo {volume} {6}},\ \bibinfo {pages} {1} (\bibinfo {year}
  {2016})}\BibitemShut {NoStop}%
\bibitem [{\citenamefont {Larremore}\ \emph {et~al.}(2014)\citenamefont
  {Larremore}, \citenamefont {Shew}, \citenamefont {Ott}, \citenamefont
  {Sorrentino},\ and\ \citenamefont {Restrepo}}]{larremore2014inhibition}%
  \BibitemOpen
  \bibfield  {author} {\bibinfo {author} {\bibfnamefont {D.~B.}\ \bibnamefont
  {Larremore}}, \bibinfo {author} {\bibfnamefont {W.~L.}\ \bibnamefont {Shew}},
  \bibinfo {author} {\bibfnamefont {E.}~\bibnamefont {Ott}}, \bibinfo {author}
  {\bibfnamefont {F.}~\bibnamefont {Sorrentino}}, \ and\ \bibinfo {author}
  {\bibfnamefont {J.~G.}\ \bibnamefont {Restrepo}},\ }\href@noop {} {\bibfield
  {journal} {\bibinfo  {journal} {Physical review letters}\ }\textbf {\bibinfo
  {volume} {112}},\ \bibinfo {pages} {138103} (\bibinfo {year}
  {2014})}\BibitemShut {NoStop}%
\bibitem [{\citenamefont {Martin}\ \emph {et~al.}(2010)\citenamefont {Martin},
  \citenamefont {Shreim},\ and\ \citenamefont {Paczuski}}]{martin2010activity}%
  \BibitemOpen
  \bibfield  {author} {\bibinfo {author} {\bibfnamefont {E.}~\bibnamefont
  {Martin}}, \bibinfo {author} {\bibfnamefont {A.}~\bibnamefont {Shreim}}, \
  and\ \bibinfo {author} {\bibfnamefont {M.}~\bibnamefont {Paczuski}},\
  }\href@noop {} {\bibfield  {journal} {\bibinfo  {journal} {Physical Review
  E}\ }\textbf {\bibinfo {volume} {81}},\ \bibinfo {pages} {016109} (\bibinfo
  {year} {2010})}\BibitemShut {NoStop}%
\bibitem [{\citenamefont {Wilson}\ and\ \citenamefont
  {Cowan}(1972)}]{wilson1972excitatory}%
  \BibitemOpen
  \bibfield  {author} {\bibinfo {author} {\bibfnamefont {H.~R.}\ \bibnamefont
  {Wilson}}\ and\ \bibinfo {author} {\bibfnamefont {J.~D.}\ \bibnamefont
  {Cowan}},\ }\href@noop {} {\bibfield  {journal} {\bibinfo  {journal}
  {Biophysical journal}\ }\textbf {\bibinfo {volume} {12}},\ \bibinfo {pages}
  {1} (\bibinfo {year} {1972})}\BibitemShut {NoStop}%
\bibitem [{\citenamefont {Benayoun}\ \emph {et~al.}(2010)\citenamefont
  {Benayoun}, \citenamefont {Cowan}, \citenamefont {van Drongelen},\ and\
  \citenamefont {Wallace}}]{benayoun2010avalanches}%
  \BibitemOpen
  \bibfield  {author} {\bibinfo {author} {\bibfnamefont {M.}~\bibnamefont
  {Benayoun}}, \bibinfo {author} {\bibfnamefont {J.~D.}\ \bibnamefont {Cowan}},
  \bibinfo {author} {\bibfnamefont {W.}~\bibnamefont {van Drongelen}}, \ and\
  \bibinfo {author} {\bibfnamefont {E.}~\bibnamefont {Wallace}},\ }\href@noop
  {} {\bibfield  {journal} {\bibinfo  {journal} {PLoS computational biology}\
  }\textbf {\bibinfo {volume} {6}},\ \bibinfo {pages} {e1000846} (\bibinfo
  {year} {2010})}\BibitemShut {NoStop}%
\bibitem [{\citenamefont {Xue}\ \emph {et~al.}(2020)\citenamefont {Xue},
  \citenamefont {Li}, \citenamefont {Grassberger},\ and\ \citenamefont
  {Chen}}]{xue2020swarming}%
  \BibitemOpen
  \bibfield  {author} {\bibinfo {author} {\bibfnamefont {T.}~\bibnamefont
  {Xue}}, \bibinfo {author} {\bibfnamefont {X.}~\bibnamefont {Li}}, \bibinfo
  {author} {\bibfnamefont {P.}~\bibnamefont {Grassberger}}, \ and\ \bibinfo
  {author} {\bibfnamefont {L.}~\bibnamefont {Chen}},\ }\href@noop {} {\bibfield
   {journal} {\bibinfo  {journal} {Physical Review Research}\ }\textbf
  {\bibinfo {volume} {2}},\ \bibinfo {pages} {042017} (\bibinfo {year}
  {2020})}\BibitemShut {NoStop}%
\end{thebibliography}%
\newpage

\setcounter{equation}{0}
\setcounter{figure}{0}
\renewcommand\thefigure{SM\arabic{figure}}
\renewcommand\theequation{SM\arabic{equation}}

\begin{center}
	
	\Large{\textbf{Supplemental Material}}
	
\end{center}

In this supplementary material, we present some details of the paper.
\section{Kinouchi-Copelli (KC) model}
\begin{figure}
	\includegraphics[width=80mm]{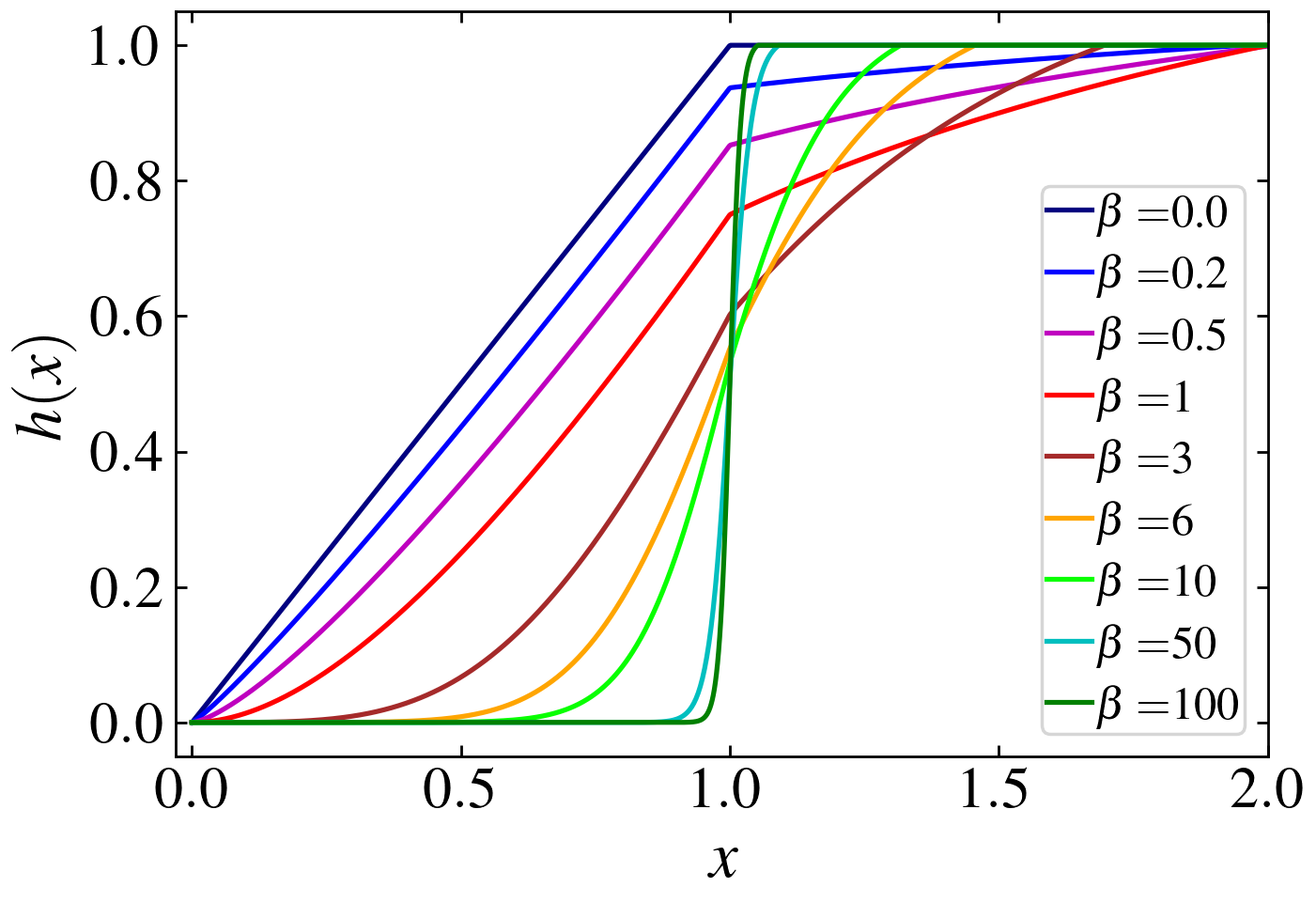}
	\caption{ $h(x)$ as an interpolation between deterministic and stochastic spike dynamics, Eq.~3 in the paper.}
	\label{fig:Interpolation}
\end{figure}

The dynamical transfer function that we use is shown in Fig.~\ref{fig:Interpolation}. It is seen that for $\beta=0$, the function reduces to the customary linear one, and by increasing $\beta$ the function changes, reaching to an extreme limit $\beta=\infty$ which is a deterministic (step-like) function.\\

To explain our model, let us start with the Kinouchi-Copelli (KC) model for a system with $N$ excitable nodes and the variable $\left\lbrace x_i\right\rbrace_{i=1}^N$ which takes two values $0,1,...,m$ in which $0$ is the rest sate, and $1$ is the excite state, and $x_i=2,...,m$ shows the refractory state. If node $x_i(t)=0$, it becomes excited in the next step ($x_i(t+1)=1$) by the neighboring excited node $j$ with probability $A_{ij}$, or independently by an external stimuli with the probability $\eta$. The nodes in the state $x_i(t+1)=x_i(t)+1$ if $1\le x_i(t)<m-1$, and $x_i(t+1)=0$ if $x_i(t)=m-1$. For $m=2$ it is not hard to show that (assuming a locally tree-like network)
\begin{equation}
p_i^{t+1}=(1-p_i^t)\left(\eta+(1-\eta)\left[1-\prod_j^N(1-p_j^tA_{ij}) \right] \right) 
\label{Eq:masterEq}
\end{equation} 
Note that the first factor guarantees that the site $i$ is in rest at time $t$. To understand the second factor, note that if $p_j^tA_{ij}=0$ for all neighbors, then $p_i^{t+1}=(1-p_i^t)\eta$ (which is due to external stimuli) and when $p_j^tA_{ij}=1$ for at least one $j$, then $p_i^{t+1}=(1-p_i^t)$, i.e. it turns on definitely if $p_i^t=0$. Then, for testing the stability of the solution $p^*=0$, one can expand the equation for small $p_i^t$s (to the first order) and in the limit of zero external stimuli the following equation is obtained:
\begin{equation}
p_i^{t+1}=(1-p_i^t)\eta +(1-\eta)\sum_j^Np_j^tA_{ij}\rightarrow \sum_j^Np_j^tA_{ij},
\label{Eq:linearizedEq}
\end{equation} 
which admits the solution $p_i^t=\lambda^tu_i$ where $u_i$ and $\lambda$ are the eigenvector and (largest) eigenvalue of $A$ matrix. The equation \ref{Eq:masterEq} holds for the case where at least one node can excite another node. Now let us consider the case where \textit{two} excited nodes are necessary for exciting \textit{one} node. Then this equation changes to:
\begin{equation}
p_i^{t+1}=(1-p_i^t)\left(\eta+(1-\eta)\left[1-\prod_{j>k}^N(1-p_j^tp_k^tA_{ij}A_{ik}) \right] \right) 
\label{Eq:masterEq2}
\end{equation}
which, for small $p$ limit and $\eta\rightarrow 0$ casts to
\begin{equation}
\begin{split}
p_i^{t+1}&=\sum_{j>k}p_j^tp_k^tA_{ij}A_{ik}\\
&= \frac{1}{2}\left[ \left( \sum_j^Np_j^tA_{ij}\right)^2-\sum_j^N\left(p_j^tA_{ij}\right)^2\right] 
\end{split}
\label{Eq:linearizedEq2}
\end{equation}
for which the first term is the leading term. For a general case where $\alpha$ excited nodes are required for exciting, we obtain 
\begin{equation}
p_i^{t+1}\rightarrow \frac{1}{\alpha!}\left( \sum_{j=1}^Np_j^tA_{ij}\right)^{\alpha}
\label{Eq:linearizedEqAlpha}
\end{equation}
which is a generalized version of KC model. Note that summing over all possible $\alpha$s gives $\exp\left[ \sum_j^Np_j^tA_{ij}\right]-1 $ which is a standard (exponential) dynamic function (a normalization factor is needed) used in a large class of excitable networks.

\section{mean field analysis}
In this section we map the Eq.~3 in the paper to a one-dimensional map using a mean-field scheme. We first consider the case $\mathbb{I}$. Suppose we excite node $i$ at time $t=0$, so that $x_n^0=\delta_{n,i}$. We will calculate the expected number of nodes that are
excited at time $t=1$. We have
\begin{equation}
x_n^1=\left\lbrace \begin{matrix}
1 & \text{with probability} \ h_{\beta}\left(\sum_{m=1}^NA_{nm}x_m^t \right), \ n\ne i \\
0 & \text{otherwise}
\end{matrix}\right. 
\end{equation}
The expected network activity $s_t\equiv \frac{1}{N}\sum_{i=1}^N x_i^t$ at $t=1$ is given by
\begin{equation}
\begin{split}
\mathbb{E}\left[s_1\right] &=\frac{1}{N}\sum_{n=1}^N\mathbb{E}\left[x_n^1\right]\\
&=\frac{1}{N}\sum_{n\ne i}^N h_{\beta}\left( \sum_{m=1}^NA_{nm}x_m^0\right),
\end{split}
\end{equation}
where $\mathbb{E}\left[s_{t}\right]$ is the ensemble average of $s_{t}$. Since $x_n^0=\delta_{n,i}$, this simplifies to
\begin{equation}
\mathbb{E}\left[s_1\right]=\frac{1}{N}\sum_{n\ne i}^N h_{\beta}\left( A_{ni}\right).
\end{equation}
Now, the adjacency matrix is weighted, and we can write $A_{nm}=a_{nm}w_{nm}$, where $a_{nm}$ is 1 (0) if nodes $n,m$ are
connected (not connected), and the weights $w_{nm}$ are randomly and independently chosen from a uniform distribution
in $[0,2\sigma]$, where $\sigma$ is chosen so that the largest eigenvalue of the matrix $A$ is $\lambda$. Since the network is Erd\H{o}s-R\'enyi
with mean degree $\left\langle k\right\rangle$, we have the relation $\left\langle k\right\rangle \sigma \approx \lambda$. Rewriting the right hand side of the previous equation as an average over nodes,
\begin{equation}
\mathbb{E}\left[s_1\right]=\frac{1}{N}\sum_{n\ne i}^Na_{ni}h_{\beta}\left(w_{ni} \right)= \langle ah_{\beta}\left(w\right)\rangle
\end{equation}
and using independence, we find
\begin{equation}
\begin{split}
\mathbb{E}\left[s_1\right]&=\langle a\rangle \langle h_{\beta}\left(w\right)\rangle\\
&=\frac{\langle k\rangle }{N} \langle h_{\beta}\left(w\right)\rangle
\end{split}
\end{equation}
Finally, since $s_0=1/N$ we obtain the branching function $b$ at $t=0$ by calculating the average over the uniform distribution of weights:
\begin{equation}
b=	\mathbb{E}\left[s_1\right]/s_0=\left\langle k\right\rangle \frac{\left\langle k\right\rangle }{2\lambda}\int_0^{2\lambda/\left\langle k\right\rangle }  h_{\beta}\left( w\right)dw.
\end{equation}
Changing variables $u=w\left\langle k\right\rangle /(2\lambda)$, gives
\begin{equation}
b=	\left\langle k\right\rangle \int_0^1  h_{\beta}\left( \frac{2u\lambda}{\left\langle k\right\rangle }\right)du.
\end{equation}
On average, we expect activity to die when $b<1$. Note that $b$ only depends on the mean degree $\left\langle k\right\rangle $ and not on $N$. However, if one fixes the probability of connection $q$ so that $\left\langle k\right\rangle=qN $, then
\begin{equation}
b=qN\int_0^1h_{\beta}\left( \frac{2u\lambda}{qN }\right)du
\end{equation}
The transition point separating the regimes where the single node ends in the absorbing or critical state is therefore a solution of
\begin{equation}
1=qN\int_0^1h_{\beta}\left( \frac{2u\lambda_c}{qN }\right)du
\label{Eq:Branching}
\end{equation}
In the thermodynamic limit $N\rightarrow\infty$ one can expand the integrand. We use the following expansion for $\epsilon\ll 1$
\begin{equation}
h_{\beta}(\epsilon)=2(1-\frac{1}{\pi}\tan^{-1}\beta)\epsilon^{1+\beta}+\ \text{higher orders of}\ \epsilon
\end{equation}
so that Eq.~\ref{Eq:Branching}, in the limit $N\rightarrow\infty$ becomes
\begin{equation}
1=\frac{2^{2+\beta}}{2+\beta} (1-\frac{1}{\pi}\tan^{-1}\beta)\left(\frac{\lambda^{1+\beta}}{q^{\beta}N^{\beta}} \right).
\end{equation}
Therefore the critical $\lambda_c$ is found to be 
\begin{equation}
\lambda_c=\left( \frac{ \left(2+\beta\right)q^{\beta} }{2^{2+\beta}(1-\frac{1}{\pi}\tan^{-1}\beta)}\right)^{\frac{1}{1+\beta}}N^{\frac{\beta}{1+\beta}}
\end{equation}
This shows that when $N\rightarrow\infty$, $\lambda_c$ diverges for non-zero $\beta$'s, so that for finite $\lambda$ values we only have the states that end up in the absorbing state. 
\begin{figure}
	\includegraphics[width=90mm]{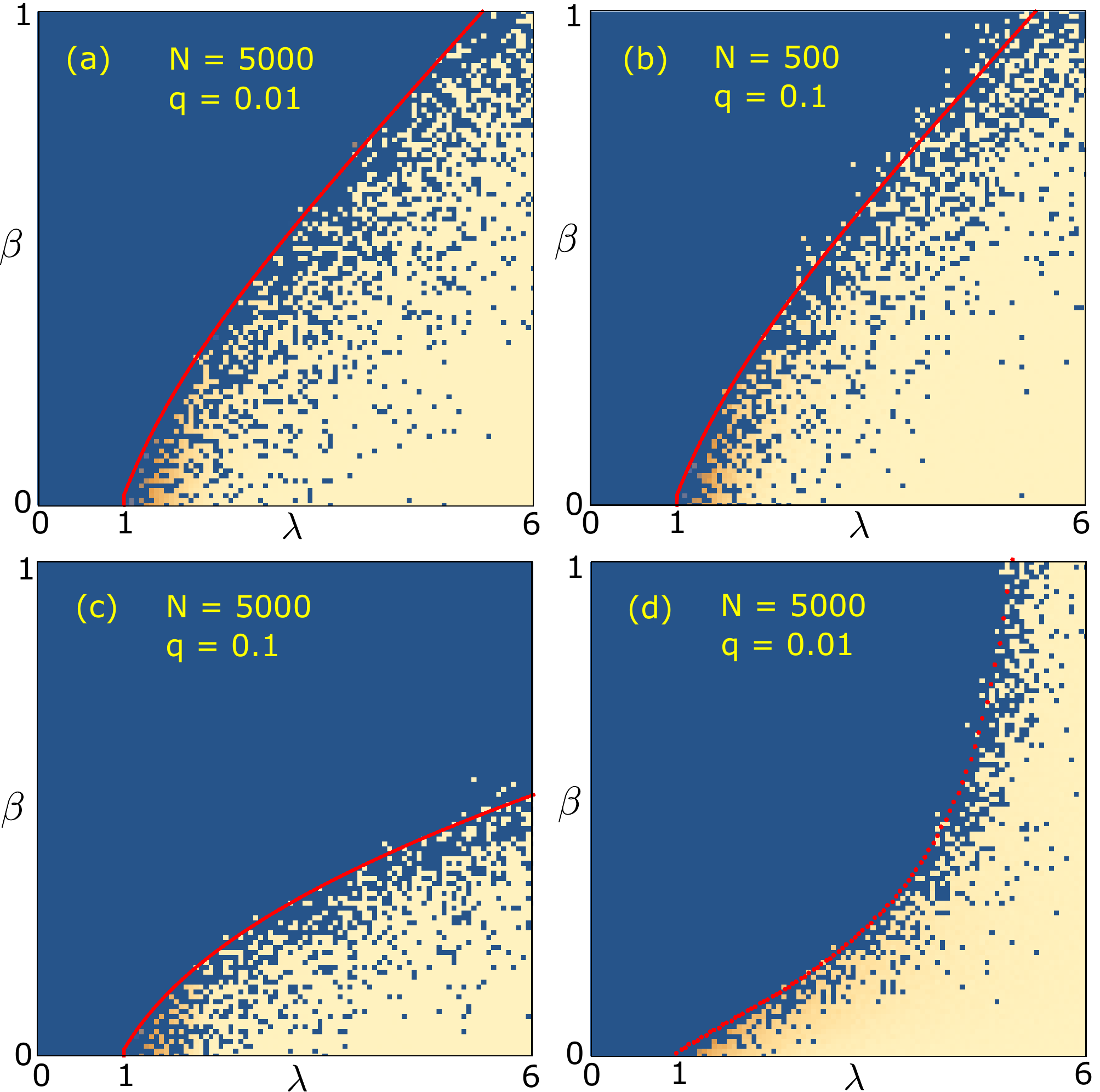}
	\caption{simulation and MF results for the phase diagram of the system for the case $\mathbb{I}$. The blue (clean) areas show systems which end up to absorbing state/super-critical states. The red line shows the criteria given by the MF analysis, i.e. Eq.~\ref{Eq:Branching}. }
	\label{fig:panels}
\end{figure}

The numerical results are shown in Fig.~\ref{fig:panels}. This figure shows the
theoretical prediction for $\lambda_c$ for $N = 5000$, $q = 0.01$, $N = 500$, $q = 0.1$, $N = 5000$, $q = 0.1$, and $N = 500$, $q = 0.01$. The red line shows the prediction of Eq.~\ref{Eq:Branching} as a function of $\lambda$. The blue dots correspond to numerical simulations of the full system that end up in the absorbing state and the clear dots to simulations that end up with positive activity. The theory predicts well the boundary between the two behaviors. For example, the two top cases have the same mean degree and behave similarly, while the behavior changes significantly when the mean degree changes (bottom panels). 
\\

Now consider the case $\mathbb{II}$, for which we use the following equation:
\begin{equation}
\begin{split}
\mathbb{E}\left[s_{t+1}\right] &=\frac{1}{N}\sum_{m=1}^N\mathbb{E}\left[x_{t+1}\right]\\
&=\frac{1}{N}\sum_{m=1}^N\mathbb{E}\left[\delta_{x_m^t,0}h_{\beta}\left( \sum_{n=1}^NA_{mn}x_n^t\right) \right]
\end{split}
\end{equation}
The sum over $N$ in the previous equation can be interpreted as an average over nodes, which we will denote with $\left\langle .\right\rangle $. Assuming independence of the random variables $\delta_{x_i^t,0}$ and $h_{\beta}\left( \sum_{j=1}^NA_{ij}x_j(t)\right) $, we get
\begin{equation}
\mathbb{E}\left[s_{t+1}\right] =\left\langle \delta_{x_i^t,0}\right\rangle  \left\langle h_{\beta}\left( \sum_{j=1}^NA_{ij}x_j(t)\right)\right\rangle 
\end{equation}
Now we use the fact that $\left\langle \delta_{x_i^t,0}\right\rangle=1-s_t $. Furthermore, for an Erd\H{o}s-R\'enyi network with large mean degree the distribution of the random variable $\sum_{m=1}^NA_{nm}x_m^t$ is narrow about its mean $\lambda s_t$, so we can approximate $\left\langle h_{\beta}\left( \sum_{j=1}^NA_{ij}x_j(t)\right)\right\rangle\simeq h_{\beta}\left(\lambda s_t \right) $, so that
\begin{equation}
s_{t+1}\equiv \mathbb{E}\left[ s_{t+1}\right]\simeq (1-s_t)h_{\beta}\left( \lambda s_t\right) 
\end{equation}
which is a one-dimensional map.\\
\\\\
\section{Hysteresis, Kurtosis and the phase diagram}
\begin{figure}[b]
	\includegraphics[width=90mm]{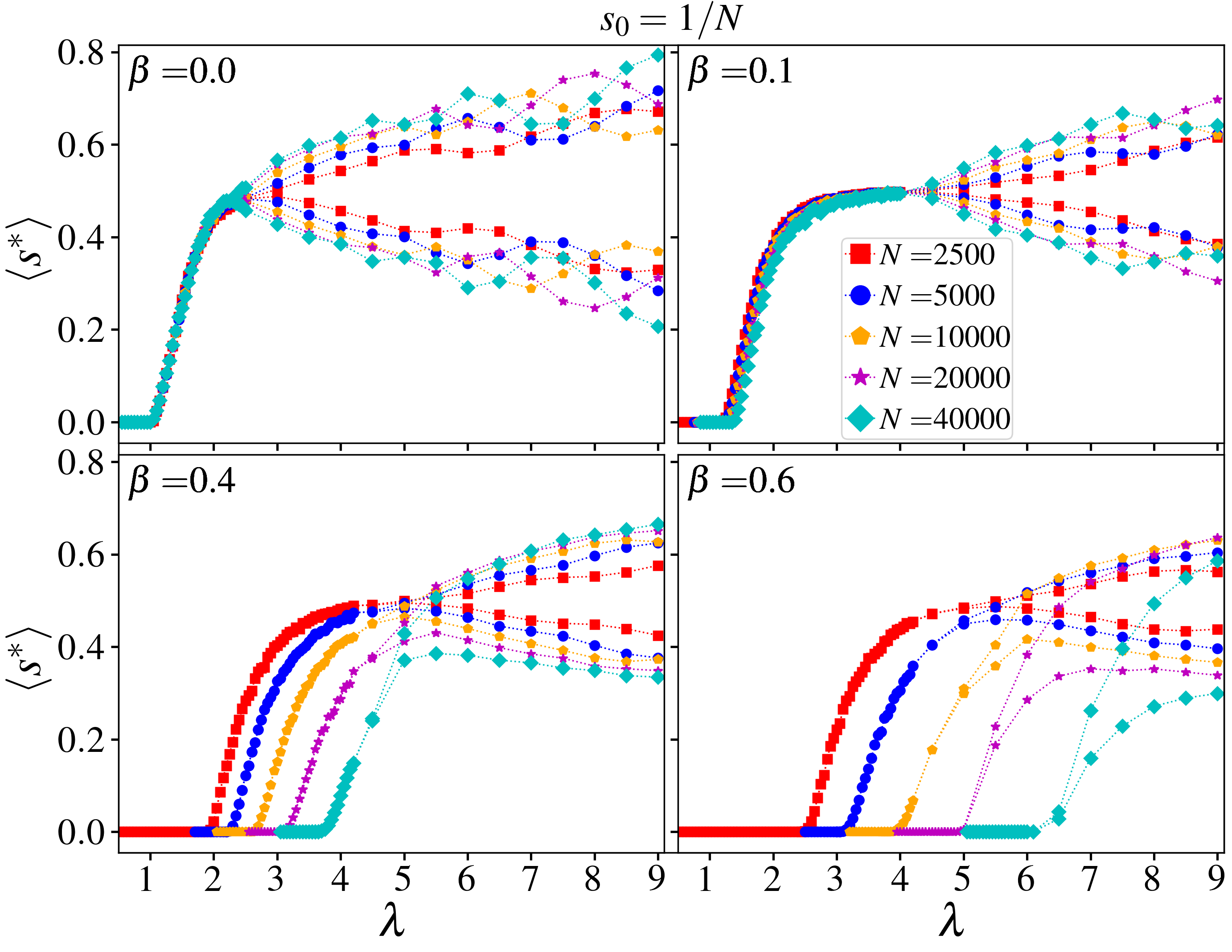}
	\caption{$\left\langle s^*\right\rangle$ in terms of $\lambda$ and $\beta$ with respect to total distribution function $P(s^*)$ showing transitions like the second order phase transition.}
	\label{fig:px-continuous}
\end{figure}

\begin{figure*}
	\includegraphics[width=85mm]{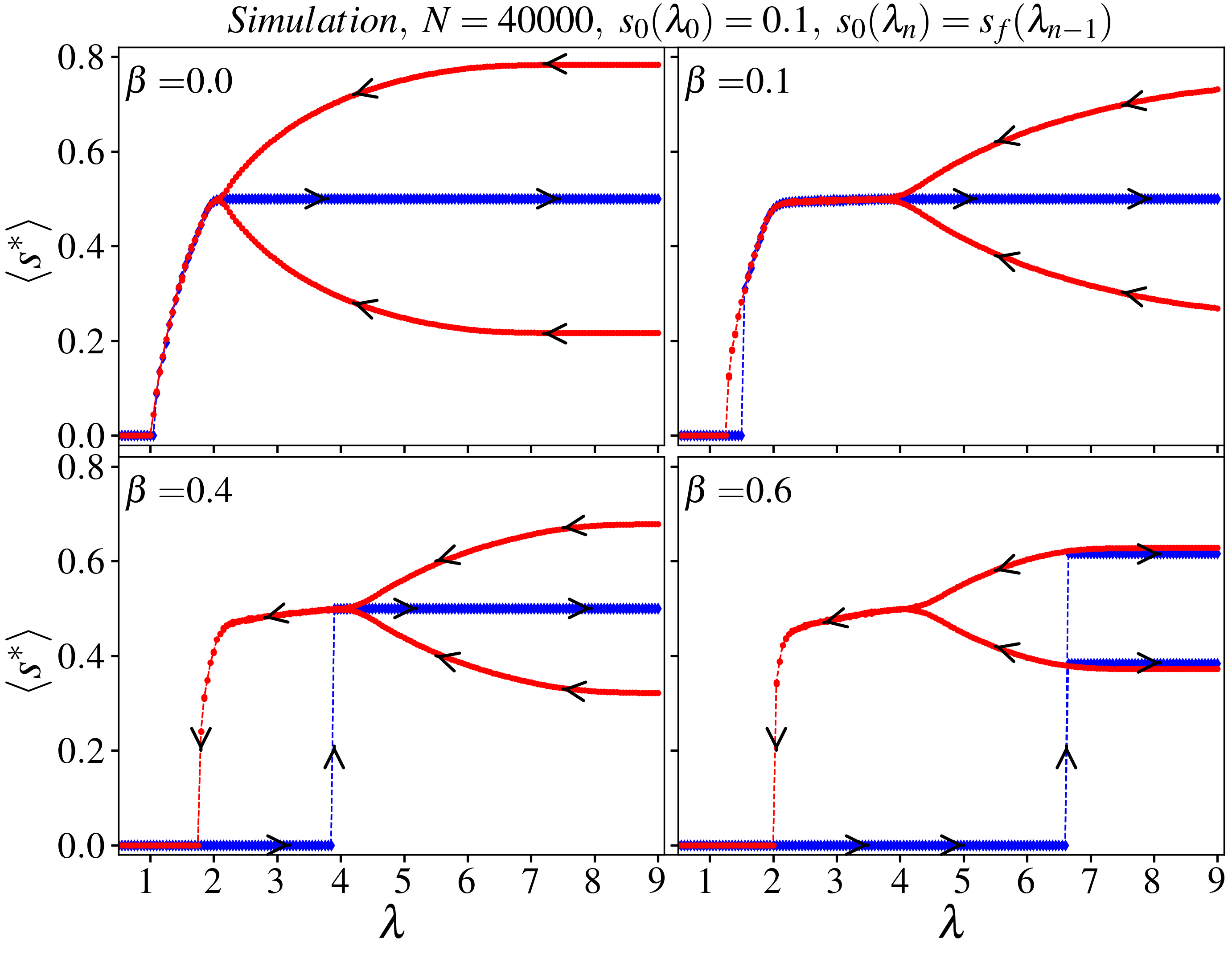}
	\includegraphics[width=85mm]{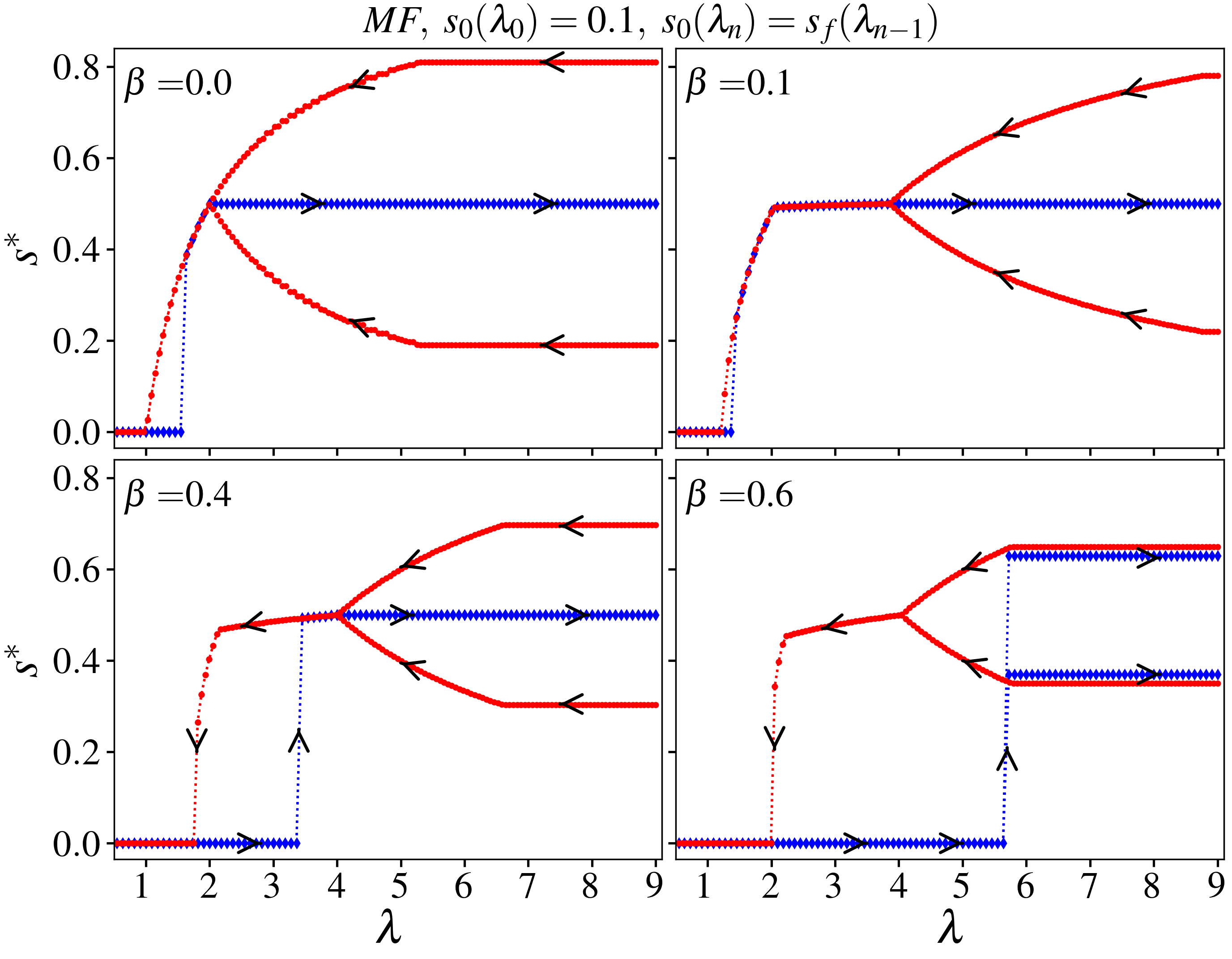}
	\caption{The simulation (left panel) and the MF (right panel) results for the hysteresis graphs for $\beta=0.4$ in terms of $N$.}
	\label{fig:hysteresis}
\end{figure*}

In this section we analyze the hysteresis and eventually we sketch the phase diagram for  case $\mathbb{I}$. First of all, we show the average $s^*$ in terms of $\lambda$ for various $\beta$ values in Fig.~\ref{fig:px-continuous} with respect to total distribution function (in the paper the averages for two branches were calculated with respect to $P_1$ and $P_2$). 
Now we consider the hysteresis effect. The corresponding graphs are shown in Fig.~\ref{fig:hysteresis} for $\beta=0.2$ for various amounts of $N$. In the left panel we show the results for the simulations, and in the right panel the MF results are shown, which are consistent with the simulation results.\\

\begin{figure}
	\includegraphics[width=90mm]{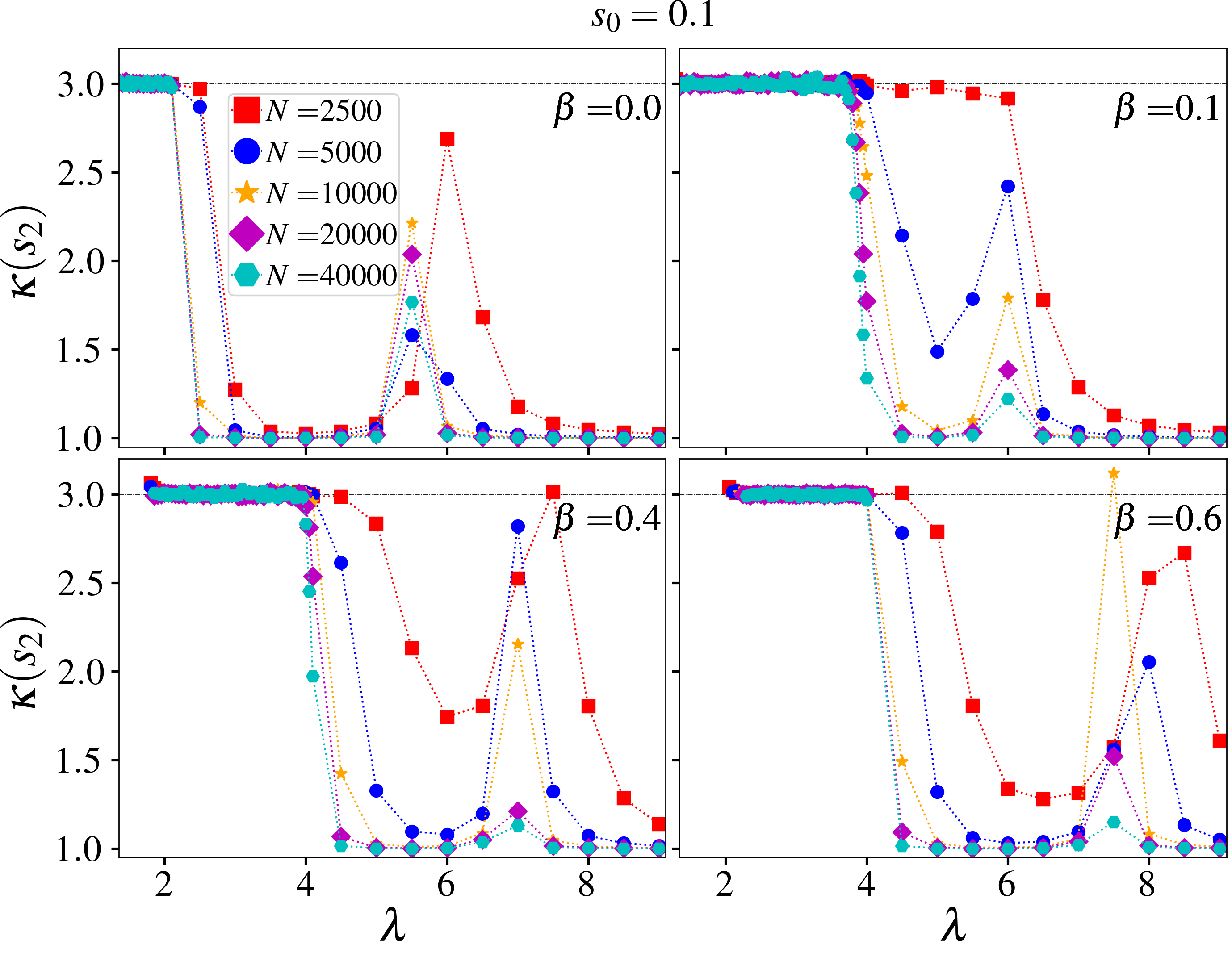}
	\caption{The Kurtosis in terms of $\lambda$ which changes behavior at $\lambda_b$.}
	\label{fig:kurtosis}
\end{figure}

The kurtosis test was employed to extract the bifurcation points, see~\cite{rahimi2021role} for details. The definition of kurtosis is as follows:
\begin{equation}
\kappa =\sigma_y^{-4}\left\langle y^4 \right\rangle
\end{equation}
Where $y=s_2-\left\langle s \right\rangle_2$, and $\sigma_y=\sqrt{\left\langle y^2\right\rangle}$. The Kurtosis is shown in Fig.~\ref{fig:kurtosis}, were $\kappa=3.0$ is a signature for the Gaussian distribution.\\

\begin{figure}
	\includegraphics[width=90mm]{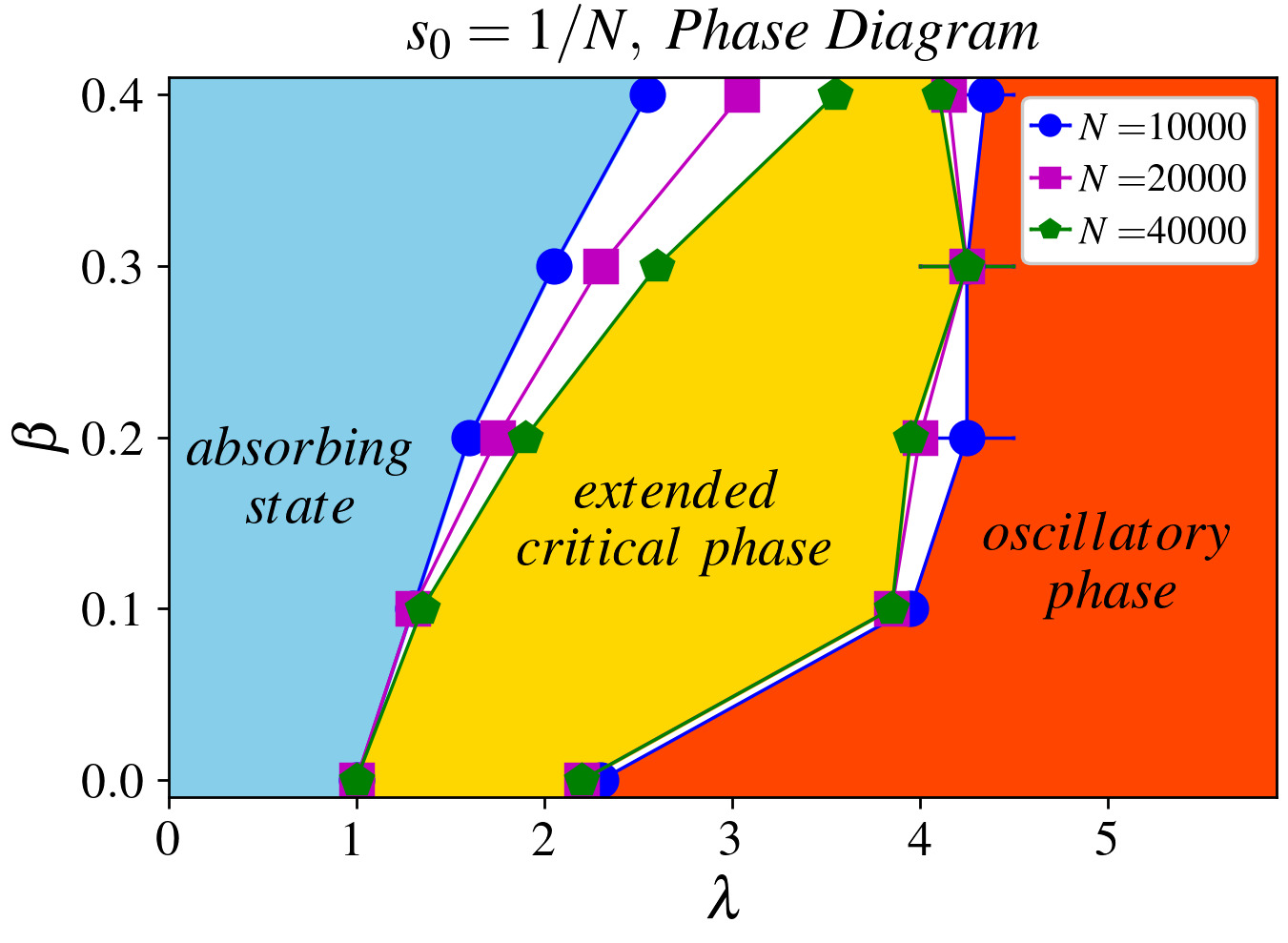}
	\caption{The phase diagram for the case $\mathbb{I}$ for three $N$ values.}
	\label{fig:PhaseDiagram2}
\end{figure}

In the Fig.~\ref{fig:PhaseDiagram2} we show the phase diagram for the case $\mathbb{I}$ for large but finite $N$'s. Remember that in the thermodynamic limit, the dominant phase for all $\lambda$ values and $\beta>0$ is the absorbing state. 

\end{document}